\documentclass[12pt, a4paper]{article}

\usepackage{amsfonts}
\usepackage{amssymb}
\usepackage{enumerate}
\usepackage{amsthm}
\usepackage{array}
\usepackage{epsfig}
\usepackage[small,bf]{caption}
\usepackage{appendix}
\usepackage{dsfont}
\usepackage{hyperref}
\usepackage{amsmath}
\usepackage{epstopdf}
\usepackage{mathrsfs}

\allowdisplaybreaks \numberwithin{equation}{section}

\begin{document}

\begin{titlepage}
 \thispagestyle{empty}

\begin{flushright}
 \end{flushright}

 \begin{center}

 \vspace{30mm}

     { \LARGE{\bf  {Fast Scramblers, Democratic Walks and Information Fields}}}

     \vspace{40pt}

\Large{{\bf Javier M. Mag\'an}} \\[8mm]
{\small\slshape
Institute for Theoretical Physics \emph{and} Center for Extreme Matter and Emergent Phenomena, \\
Utrecht University, 3508 TD Utrecht, The Netherlands \\

\vspace{5mm}

{\upshape\ttfamily j.martinezmagan@uu.nl}\\[3mm]}

\vspace{8mm}

     \vspace{10pt}

    \vspace{10pt}

\date{\today}

\end{center}

\begin{abstract}
We study a family of weighted random walks on complete graphs. These `democratic walks' turn out to be explicitly solvable, and we find the hierarchy window for which the characteristic time scale saturates the so-called fast scrambling conjecture. We show that these democratic walks describe well the properties of information spreading in systems in which every degree of freedom interacts with every other degree of freedom, such as Matrix or infinite range models. The argument is based on the analysis of suitably defined `Information fields' ($\mathcal{I}$), which are shown to evolve stochastically towards stationarity due to unitarity of the microscopic model. The model implies that in democratic systems, stabilization of one subsystem is equivalent to global scrambling. We use these results to study scrambling of infalling perturbations in black hole backgrounds, and argue that the near horizon running coupling constants are connected to entanglement evolution of single particle perturbations in democratic systems.
\end{abstract}
 \vspace{10pt}
\noindent

\end{titlepage}

\thispagestyle{plain}

\tableofcontents

\baselineskip 6 mm

\newpage

\section{Introduction}\label{sec1}

In this article we make some steps towards a better understanding of information spreading in black hole physics. When considering local and translationally invariant systems, certain aspects of information spreading can be modeled with diffusive Fokker-Plank type equations, since these equations govern the distribution of locally conserved charges in the system. On the other hand, in the context of black hole physics it is not even clear that an analogue of such simple equations might hold, since the theories that are expected to describe these enigmatic objects are theories in which every degree of freedom interacts with every other degree of freedom, such as matrix models or AdS/CFT \cite{matrix,adscft,susskind}. These matrix models show interaction terms of the type
\begin{equation}\label{nonlocal}
{\textrm {Tr}}\, M^4 = \sum_{ijln} \, M_{ij}M_{jl}M_{ln}M_{ni} \;,
\end{equation}
which are completely non-local, connecting every degree of freedom with every other democratically \footnote{See \cite{subir} for a recent construction supporting democratic systems as good microscopic models of black holes.}. In the case of AdS/CFT, terms like~(\ref{nonlocal}) govern the internal dynamics of a thermal cell at finite temperature $\beta$ \footnote{By a thermal cell we mean a spatial domain of size $\beta$ in the field theory side. It has $\mathcal{O} (c)$  degrees of freedom,  where $c$ is the central charge of the dual field theory, since the entropy is given on general grounds by $S_{\textrm{CFT}}\propto c \, (V/\beta)^{d}$. See \cite{us1,us2} for a discussion of this point.}, as argued in generic terms in \cite{susskind}. In some special cases one can explicitly show \cite{us2} that the internal dynamics of the thermal cell is given by the BFSS matrix model presented in \cite{matrix}. Throughout this article we will call these type of models { \it democratic models}. It is plain that in these models there are no natural diffusion equations of conserved charges over the different degrees of freedom. Indeed, there is even no immediate notion of space whatsoever, and the question of information spreading through the system becomes fuzzier.

As a first breakthrough over these kind of questions, based on previous work \cite{hayden} and with input from the fields of string theory and quantum information, it was conjectured in \cite{susskind} that such systems scramble information as fast as possible, with a time scale of order
\begin{equation}\label{scalesc}
t_{\textrm{scrambling}}\sim \beta \log N \;,
\end{equation}
where $\beta$ is the temperature of the system and $N$ is the number of degrees of freedom in the matrix model or in the thermal cell in the case of AdS/CFT.

This conjecture has become an extremely useful guiding light into the dynamics of black holes, since information scrambling/spreading is ultimately connected with the microscopic structure of interactions of the theory. More concretely it raises several interesting questions, of which we would like to highlight two. 

The first question is methodological, and refers to how we characterize the spread of information in the system. In a classical setting, in \cite{susskind} it was characterized as the spreading of induced charges at the so-called brick wall \cite{brick} or  streched horizon\cite{complementarity}, in \cite{us1} it was characterized as a Lyapunov time of a strongly chaotic system,  and in \cite{us3,us4} diffusion processes, or random walks over the set of degrees of freedom were used in this regard. In more rigorous quantum setups, commutators between different operators in the system at different times were used to characterize signal propagation \cite{lashkari,us3, stanford2, stanford3, verlinde, bound}, while the relaxation of reduced subsystems and entanglement entropies were studied in \cite{pdesings,lashkari,us3,stanford2014,sahakian,sahakian2,cesarscrambling,uscod,disrupting,sekino,joansimon}.

The second question is dynamical, and looks for the different classes of theories showing fast scrambling behavior, in the sense of producing the scale~(\ref{scalesc}) within some characterization of information spreading. In this regard some examples of non-local systems have been studied in \cite{pdesings,berenstein,matrixnum,lashkari,fischler,us4,stanford2,stanford3,sahakian,sahakian2,verlinde,bound,joansimon}. On the other hand, in \cite{us1} and specially in \cite{us3} it was noticed that non-locality is not a necessary condition for fast scrambling by showing that local theories on hyperbolic spaces or expander graphs \cite{avi1} are natural examples of fast scramblers. Lastly, the chaotic properties of discretizations of near horizon AdS geometries have been studied in \cite{modular}.

Given the previous context and questions, the objective of this article is twofold. From the dynamical perspective, and within the arguably simplest characterization of information spreading, which is that of a Markov process or random walk over the degrees of freedom of the system\footnote{See section~(\ref{sec2}) for a generic description of these type of dynamics.}, we first want to ask for the families of Markovian kernels with fast scrambling behavior.

In Ref \cite{us3,us4} two such families were presented. The first family corresponds to random walks defined on expander graphs \cite{avi1,us3}, or their continuum analogues given by diffusion processes on hyperbolic spaces. In both cases the spectral gap of the Markov transition matrix is bounded away from zero in the thermodynamic limit. As reviewed in section~(\ref{sec2}), this is a sufficient condition to relax in a time of $\mathcal{O}(\log N)$. The importance of these systems in the black hole context was described in \cite{us1,us3} by showing that the near horizon region of thermal event horizons is naturally described by a homogeneous thermal ensemble in a hyperboloid. The second family was described in \cite{us4}, and it is inspired by the first. It corresponds to random walks defined on ultrametric graphs. These are weighted graphs in which distances show an ultrametric structure (see \cite{ultrametricity} for a description of ultrametricity). In this case the gap is not bounded away from zero, but the second eigenvalue $\lambda$ is of order $\mathcal{O}(1/\log N)$, naturally generating the fast scrambling time scale. In this case, the connection with black hole physics lies on the natural ultrametricity generated at the event horizon due to the near horizon hyperbolicity, see \cite{us4}.

But to connect with non-perturbative formulations \cite{matrix,adscft}, and given the completely non-local interaction structure of~(\ref{nonlocal}), we want to ask if there are also fully non-local walks with fast scrambling behavior. We will refer to such non-local walks as {\it democratic walks.} The answer comes out positively. Moreover, we find a class of democratic walks with properties akin to the walks on expander graphs, i.e with non-vanishing spectral gap, and also a class of democratic walks with properties akin to the walks on ultrametric spaces. The structure of information spreading of these models will be highly simple and peculiar, as we describe in various places through the article, and it differs from any previously considered one. In particular it differs from the one proposed in \cite{susskind}, based on the analysis of induced charge spreading at the streched horizon.

The second objective of this article is to generically connect these classical random walks with microscopic models of quantum thermalization. As we mentioned before, when modeling information spreading with a random walk, we always have in mind that the underlying system possesses some locally conserved charge to which one can attach a density distribution. This density distribution (after several approximations) is argued to follow a simple Markovian equation, like a Fokker-Plank type diffusion equation. But in many-body quantum mechanics it is not always simple to identify such locally conserved charges. On the other hand, an expected \emph{information spreading} due to microscopic interactions, together with its conservation due to unitarity of the evolution, suggests the existence of some kind of \emph{information density} over the set of degrees of freedom. In this article we construct examples of such densities\footnote{Although the construction is rigorous and unambigous, it will be left unclear wheather the definition is the most appropriate one. It is certainly possible that other definitions of conceptually similar information densities exist. For the present being we are just interested in showing the existence of these densities, and use them to understand information spreading and thermalization in many body quantum mechanics in terms of classical stochastic processes.}, which we term {\it information fields} $\mathcal{I}_{\textbf{A}}^{n}(\rho)$, associated to some subsystem 
$\textbf{A}$ and microscopic state $\rho$, where $n=1,\cdots ,N$ runs over the different degrees of freedom of the system. The construction is based on previous developments in the context of quantum thermalization \cite{uscod}. We show that these information fields are defined unambigously from the quantum state, and therefore the unitary evolution of $\rho$, given by $\rho (t)=U(t)\,\rho\, U^{\dagger}(t)$, induces a well defined time evolution for $\mathcal{I}_{\textbf{A}}^{n}(\rho)$, given by $\mathcal{I}_{\textbf{A}}^{n}(\rho (t))$. This induced evolution is shown to be stochastic (it conserves the normalization of the field, $\sum_{n}\mathcal{I}_{\textbf{A}}^{n}(\rho (t))=1$) if and only if the underlying evolution is unitary. Finally, it is shown that the \emph{typical} information field, or the information field in a random state, is just the uniform distribution $\mathcal{I}_{\textbf{A}}^{n}(\rho_{\textrm{random}})=1/N$. Therefore, any non-equilibrium thermal process will drive stochastically the information field towards stationarity in the usual sense. This then provides a novel and interesting framework to study information spreading in many-body quantum systems in terms of classical stochastic processes. In particular it is potentially a way to study quantum chaotic phenomena by the analysis of the classical mixing properties of these information densities over their phase space, which is just the set of degrees of freedom. It also provides a novel perspective of the fast scrambling conjecture \cite{susskind}. From this perspective, the conjecture puts constraints on the possible stochastic evolutions of the information field $\mathcal{I}_{\textbf{A}}(\rho)$.

In section~(\ref{sec4}) we investigate the previous construction, regarding quantum thermalization and information fields, in the context of democratic models. With or without assuming democratic walks as the stochastic evolution kernels, the structure of information spreading will be seen to be different form the one advocated in \cite{susskind}, regarding the spreading of conserved charges at the streched horizon \footnote{There is no contradiction here, since the process described in \cite{susskind} is not really a scrambling process at the streched horizon, being the classical imprint of a freely falling particle through the near horizon region to the streched horizon, see \cite{us1}.}. It will also differ from the structure of information spreading in random circuit models, as described in section~(\ref{nonv}). More concretely we will show that information spreading in non-local systems is instantaneous, and therefore the global properties of information spreading, and its characteristic time scales, such as the scrambling time, are seen already at the level of local relaxation of perturbations. This will be clear at the level of entanglement entropies, for which we will argue that any subsystem, no matter the size, equilibrates at the same time. In this precise sense, we will conclude that in democratic systems, {\it global relaxation/scrambling} is equivalent to {\it local relaxation}. Intuitively, given the stabilization of one subsystem, for example measured by stabilization of its entanglement entropy, unitarity implies that the lost information has to be found somewhere in the rest of the system. But democracy of interactions implies that it has to be equidistributed over all degrees of freedom, i.e democracy implies that the lost information is instantaneously scrambled. Scrambling then just amounts to the stabilization of the given subsystem.

In the last part of section~(\ref{sec4}) we show that a specific class of random walks qulitatively reproduce the results presented in \cite{lashkari,stanford2014,sahakian2,joansimon}, together with the black hole estimate done in section~(\ref{sec5}), while still giving a coherent view of global information spreading. Also, given the connection between unitarity and stochasticity of the present construction, we consider the thermodynamic limit of the democratic walk. This thermodynamic limit is highly peculiar and it is very different from the thermodynamic limit of local walks, such as those defined on euclidean or expander graphs. The analysis will shed new light on the fate of information in black hole phyiscs, and more concretely on the issue of information loss.

Finally, the last section is devoted to use the gained insights in the context of black holes, dually described by democratic models, as commented before. It is argued that by the time an infalling peturbation reaches the brick wall/streched horizon, the information associated to it is fully scrambled over the color degrees of freedom of the matrix model. The argument connects the near horizon running coupling constants with the growth of entanglement entropy of the single infalling perturbation. This growth is suppressed at early times, since the couplings are suppressed by powers of the Planck mass. By the time the perturbation reaches the streched horizon the couplings become of $\mathcal{O}(1)$, and the perturbation becomes fully entangled. If the underlying interactions are democratic, the shared entanglement is democratically spread over the whole set of degrees of freedom, meaning that the system has already been globally scrambled. The argument then relies deeply on the previously described equivalence between global scrambling and local relaxation, valid for democratic systems.

\section{Democratic walks}\label{sec2}

As described in the introduction, understanding generic aspects of democratic models seems crucial for understanding aspects of black hole dynamics. With this in mind we define and study a generic class of random walks which are characterized by being defined on complete graphs. To our knowledge they constitute a new family of Markov processes and they may be of interest on their own. Throughout the article we will refer to them as {\it democratic walks}. Before focusing on this particular new class, let us describe some generic aspects of random walks.

The problem of a random walk in an abstract graph is a famous and recurring problem in both mathematics and physics. It is concerned with the dynamics of an abstract object which can be in $N$ different states. The statistical description at a given instant of time is given by means of a probability distribution. The probability distribution provides the probabilities of finding the object in each of the $N$ states, which can be thought as $N$ `locations' in an abstract space. Therefore, we have a vector $p_{i}$ of $N$ components which must satisfy the probability constraint given by 
$$\sum\limits_{i=1}^{N}p_{i}=1\;.$$
The dynamics proceeds by means of a Markov process. This is defined as a linear relationship which yields the probability distribution at step $n+1$ from the probability distribution at step $n$ \footnote{One could use continuous time with a related equation of the type $\partial_{t}\rho_{i}(t)=\bar{M}_{ij}\rho_{j}(t)$. The results would not change and we will take the discrete time description.}. More concretely we have
\begin{equation}\label{walk}
p_{j}^{(n+1)}= \sum\limits_{i=1}^{N}M_{ji}\,p_{i}^{(n)} \;,
\end{equation}
where the transition matrix $M_{ij}$ is {\it row-stochastic} which implies
\begin{equation}\label{stoc}
 \sum\limits_{j=1}^{N}M_{ji}=1\;,
\end{equation}
a condition which is imposed to ensure that the evolution conserves the probability constraint. In addition,
  \begin{equation}\label{detailed}
  M_{ji}=M_{ij}\;
  \end{equation}
due to microscopic reversibility. This condition is a special case of the more generic detailed balance which occurs when the degeneracy of each possible state equals one. The physical picture is that of a particle moving on a graph\footnote{A graph is collection of \emph{vertices} and a collection of \emph{edges} between them. The graph can also be \emph{weighted}, in the sense that we can give different \emph{weights} to different edges.}, from one vertex to another vertex with probabilities given by the entries of the matrix $M$.

Exactly solving the random walk~(\ref{walk}) amounts to finding a closed expression for $M^{n}$, since then we can find the exact time evolution of any initial probability distribution by simple matrix multiplication. In the case that the exponentiation turns out to be too difficult, we can still understand several properties of the walk just by knowing certain properties of the eigenvalues of $M$.

One such property is the relaxation time of the walk, i.e the time after which the particle is at any vertex with equal probability, so that $p^{n}\sim p_{u}=1/N$. Notice that this uniform distribution $p_{u}$ is an eigenvector of $M$ with unit eigenvalue by construction. The row-stochastic nature of the transition matrix implies (via the Perron--Frobenius theorem) that the absolute value of any of the other eigenvalues is less than one, an essential ingredient for the attainment of equilibrium. With this in mind, we consider, without loss of generality, that $1\geq|\lambda_{1}|\geq|\lambda_{2}|\geq ...\geq|\lambda_{N-1}|$, where $\lambda_{i}$ are the rest of the eigenvalues of $M$. To obtain the relaxation time of the walk, notice that the knowledge of $\lambda_{1}\equiv \lambda$ already allows us to bound the distance of the evolved probability distribution from the uniform one:
\begin{equation}\label{distance}
|p^{n}-p_{u}|=|M^{n}(p^{0}-p_{u})|\leq \lambda^{n}|(p^{0}-p_{u})|\leq \lambda^{n}=(1-\Delta)^{n} \;,
\end{equation}
where $\vert p\vert =\sqrt{\sum_{i}p_{i}^{2}}$, see \cite{avi1} for a more detailed treatment. We conclude that the figure of merit in this problem is the gap of the transition matrix $\Delta \equiv 1-\lambda$, controlling the rate of approach to equilibrium of the probability distributions.

Given these preliminaries, the critical question is for the specific structure of the kernel $M_{ij}$. More intuitively, we are interested in the following question: what type of Markov kernels might encode the dynamics of information spreading in black holes?

To approach this question it is interesting and useful to take the so-called fast scrambling conjecture \cite{susskind} as a guiding light. In this setting the question gets a little bit more precise: what type of Markov kernels show some short of fast scrambling behavior \cite{susskind}? To sharpen even more the question, and with the objective of getting nearer to non-perturbative descriptions of black hole dynamics such as matrix models \cite{matrix} and AdS/CFT \cite{adscft}, we want to consider completely non-local or {\it democratic walks}, which we define below. With these two physically motivated inputs, the previous generic and opaque question turns into a concrete mathematical problem. Can we find democratic walks with characteristic time scales of $\mathcal{O}(\log N)$?

The immediate answer to this question seems negative, the argument being that by choosing a transition matrix $M$ with entries given by $M_{ij}\sim \mathcal{O}(1/N)$, we obtain a family of trivial Markov processes, in which any initial distribution relaxes/becomes uniform after one step \footnote{To observe this fact the simplest option is to choose $M_{ij}=1/N$ and plug it into~(\ref{walk}).}. But indeed, as we argue more properly in section~(\ref{sec4}), we do not expect the kernel describing information spreading in black hole physics to have non-diagonal entries of the same order as diagonal ones.

\begin{figure}[htb]
   \centering
     \includegraphics[width=0.7 \textwidth]{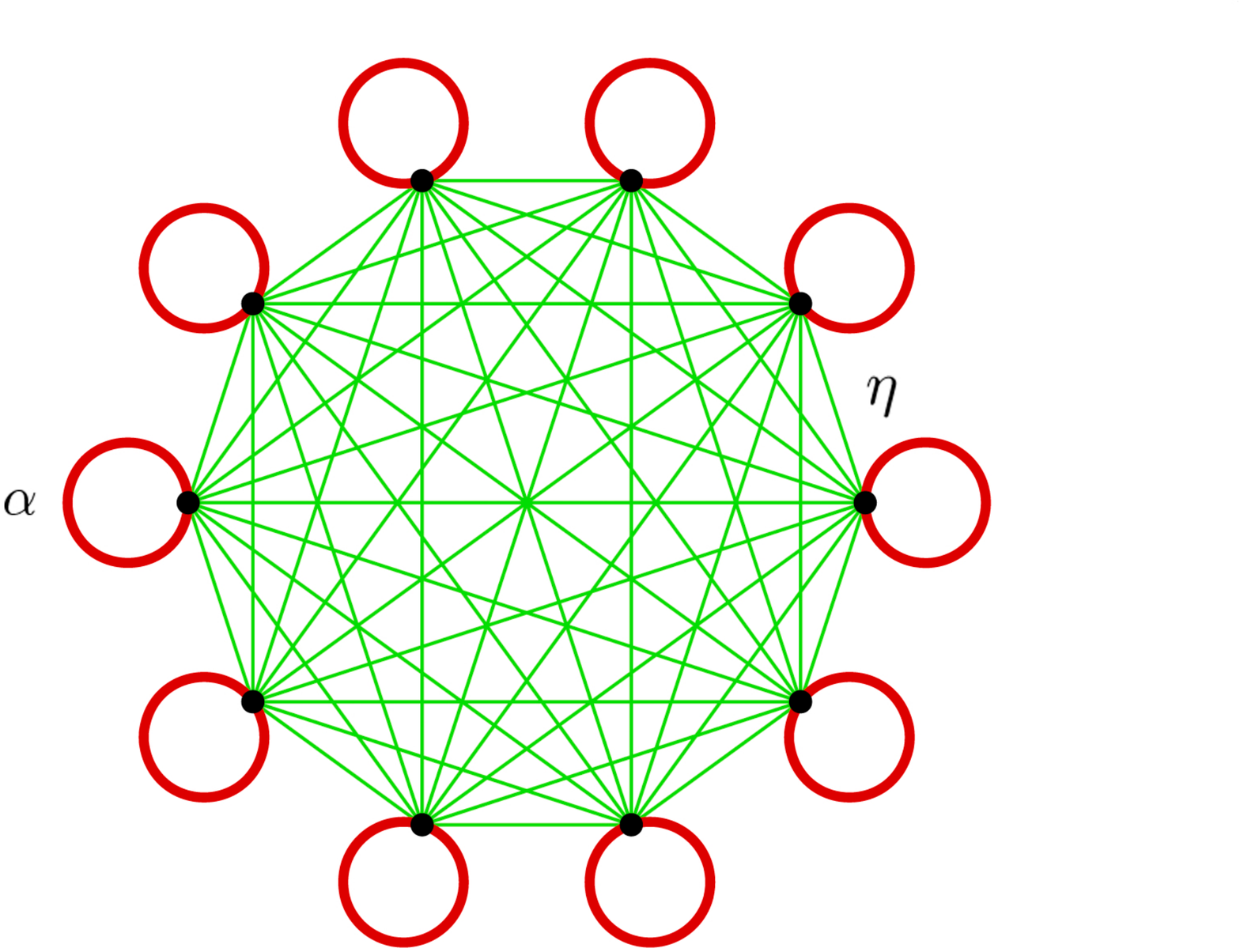}
   \caption{ Weighted graph representing the democratic walk defined by relations~(\ref{rel1}) and~(\ref{rel2}). The real number $\alpha$ corresponds to the probability of staying in the same vertex, while $\eta$ is the probability of jumping to any of the other vertices. Within this fixed weigthed graph, the random walk is explicitly solvable, the solution given by relations~(\ref{alet}) and~(\ref{solution1}).}
  \label{demo}
   
 \end{figure}

We thus define democratic walks more generically by the following relations:
\begin{equation}\label{rel1}
M_{ij}=\eta \;, \qquad \text{for} \,\,\, i \neq j
\end{equation}
and
\begin{equation}\label{rel2}
M_{ii}=\alpha=1-(N-1)\, \eta\;,
\end{equation}
which by construction satisfies the normalization constraint $\sum_{i}M_{ij}=1$. This can be seen as a random walk on a weighted complete graph, see Fig~(\ref{demo}). We fix the entries in such a way in order to have an exactly solvable model. This will be very helpful for understanding generic properties of information spreading in democratic models. Later on we will show what hierarchy of $\eta$ and $\alpha$ produces fast scrambling behavior, and we will also show that perturbing the transition matrix entries does not affect the main structure and results. But for now, let us solve the model defined by~(\ref{rel1}) and~(\ref{rel2}) generically.

We will begin with the spectral properties of $M$. For the present matrix $M$ we can apply Sylvester's determinant theorem to find the characteristic equation of the eigenvalues $\lambda$, which is given by:
\begin{equation}
(\alpha-\eta-\lambda)^{N}(1+N\frac{\eta}{\alpha-\eta-\lambda})=(\alpha-\eta-\lambda)^{N-1}(\alpha-\eta-\lambda+N\eta)=0\;.
\end{equation}
This provides a non-degenerate eigenvalue $\lambda_{u}=1$, corresponding to the uniform distribution $p_{u}=1/N$, and a $N-1$ degenerate eigenvalue given by
\begin{equation}\label{eigen}
\lambda=\alpha-\eta\;.
\end{equation}
To extract the scaling of the relaxation time with the number of sites $N$ we need to give specific values to $\alpha$ and $\eta$. We leave this for the next section, since within the fixed assumptions~(\ref{rel1}) and (\ref{rel2}) we can go further and solve the model exactly. Solving the model amounts to finding a closed expression for $M^{n}$. To compute $M^{n}$, we notice the following two statements:
\begin{itemize}
 \item $M^{n}$ has the same structure as $M$, i.e $M_{ii}=M_{jj}$ and $M_{i\neq j}=M_{k\neq l}$. We will refer to the off-diagonal elements of $M^n$ as $\eta_{n}$ and the diagonal ones as $\alpha_{n}$.
 \item $\alpha_{n}$ and $\eta_{n}$ satisfy a coupled recursion relation, given by:
\begin{equation}
\begin{pmatrix}
\alpha_{n} \\ \eta_{n}  
\end{pmatrix} =\begin{pmatrix}
\alpha & (N-1)\,\eta \\ \eta & (N-2)\,\eta+\alpha  
\end{pmatrix}\begin{pmatrix}
\alpha_{n-1} \\ \eta_{n-1}  
\end{pmatrix}=\begin{pmatrix}
\alpha & (N-1)\,\eta \\ \eta & (N-2)\,\eta+\alpha  
\end{pmatrix}^{n-1}\begin{pmatrix}
\alpha \\ \eta  
\end{pmatrix}\;.
\end{equation}
\end{itemize}
We solve the recursion relation in two steps. First notice that
\begin{equation}
\begin{pmatrix}
\alpha & (N-1)\,\eta \\ \eta & (N-2)\,\eta+\alpha  
\end{pmatrix}^{n-1}=\begin{pmatrix}
\frac{1}{\sqrt{2}} & \frac{1}{\sqrt{2}} \\ \frac{1}{\sqrt{2}} & -\frac{1}{\sqrt{2}}  
\end{pmatrix}\begin{pmatrix}
1 & -(N-2)\,\eta \\ 0 & \alpha-\eta  
\end{pmatrix}^{n-1}\begin{pmatrix}
\frac{1}{\sqrt{2}} & \frac{1}{\sqrt{2}} \\ \frac{1}{\sqrt{2}} & -\frac{1}{\sqrt{2}}  
\end{pmatrix}\;.
\end{equation}
The exponentiation left to be done boils down to a single recursion relation for the off diagonal non-vanishing term, which reads:
\begin{equation}
x_{n}=x_{n-1}-(N-2)\,\eta\, (\alpha-\eta)^{n-1}\;,
\end{equation}
with the initial condition $x_{1}=-(N-2)\,\eta$. The solution is given by
\begin{equation}
x_{n-1}=\frac{(N-2)\,\eta\, [(\alpha-\eta)^{n-1}-1]}{1-\alpha+\eta}\;,
\end{equation}
so that
\begin{equation}
\begin{pmatrix}
\alpha & (N-1)\,\eta \\ \eta & (N-2)\,\eta+\alpha  
\end{pmatrix}^{n-1}=\begin{pmatrix}
\frac{1}{\sqrt{2}} & \frac{1}{\sqrt{2}} \\ \frac{1}{\sqrt{2}} & -\frac{1}{\sqrt{2}}  
\end{pmatrix}\begin{pmatrix}
1 & x_{n-1} \\ 0 & (\alpha-\eta)^{n-1}  
\end{pmatrix}\begin{pmatrix}
\frac{1}{\sqrt{2}} & \frac{1}{\sqrt{2}} \\ \frac{1}{\sqrt{2}} & -\frac{1}{\sqrt{2}}  
\end{pmatrix}\;.
\end{equation}
Doing the matrix products we finally arrive to:
\begin{equation}\label{alet}
\begin{pmatrix}
\alpha_{n} \\ \eta_{n}  
\end{pmatrix} =\begin{pmatrix}
\alpha & (N-1)\,\eta \\ \eta & (N-2)\,\eta+\alpha  
\end{pmatrix}^{n-1}\begin{pmatrix}
\alpha \\ \eta  
\end{pmatrix}=\begin{pmatrix}
\frac{1}{N}+\frac{(N-1)}{N}(\alpha-\eta)^{n} \\   \frac{1}{N}-\frac{(\alpha-\eta)^{n}}{N}
\end{pmatrix}\;.
\end{equation}
Notice that $\alpha_{n}+(N-1)\,\eta_{n}=1$ as required by probability conservation.

Given the previous result, if we begin with a probability distribution localized in the first site $p_{i}^{0}=\delta_{i1}$, stochastic evolution results in:
\begin{equation}\label{solution1}
p^{n}=M^{n}p^{0}=M^{n}\begin{pmatrix}
1 \\ 0 \\ 0 \\ \vdots \\ 0
\end{pmatrix} = \begin{pmatrix}
\alpha_{n} \\ \eta_{n} \\ \eta_{n} \\ \vdots \\ \eta_{n}
\end{pmatrix}
\;,
\end{equation}
where the $\alpha_{n}$ and $\eta_{n}$ are defined by equation~(\ref{alet}) in terms of the initially given $\eta$ and $\alpha$. Having the exact full solution~(\ref{solution1}) teaches us various insightful and generic features of democratic walks, which are not dependent on the specific hierarchy we impose between $\alpha$ and $\eta$, but to the specific structure of the democratic kernel given by (\ref{rel1}) and (\ref{rel2}). The first important aspect is:
\begin{itemize}
\item The probability distribution spreads democratically over all the sites (different from the initially perturbed one) at any time.
\end{itemize}
This is very different from random walks in local lattices, as we discuss later. In such cases the probability distribution is only democratically spread after the relaxation time has elapsed. In this way we arrive to one of the main points of the article:
\begin{itemize}
\item For democratic models, \emph{global} relaxation is equivalent to \emph{local} relaxation.
\end{itemize}
This means that, given that any amount of information leaking from the initially non-zero entry is democratically distributed over the other sites, we just need to agree on how much we want the first entry to decay. This might indeed depend on the quantity studied, as we comment later.

\subsection{Fast scrambling democratic walks}\label{fast}

Defining $0<\lambda_{\textrm{eff}}<1$, the first interesting class of random walks is the following:
\begin{equation}
\eta=\frac{\lambda_{\textrm{eff}}}{N}\sim\mathcal{O}(1/N)\;,
\end{equation}
and therefore
\begin{equation}
\alpha=1-(N-1)\,\eta=1-(N-1)\,\frac{\lambda_{\textrm{eff}}}{N}\;.
\end{equation}
Using~(\ref{eigen}), in the thermodynamic limit the second eigenvalue $\lambda=\alpha-\eta\rightarrow\alpha$, since $\eta\rightarrow 0$ in this limit. Noticing also that $0<\alpha<1$ in the thermodynamic limit, so we conclude that the second eigenvalue is bounded away from the maximal eigenvalue $1$, but still non-zero.

We thus arrive to one of the main results of the article. Since $0<\alpha<1$, the previous democratic walk shows a non-vanishing gap in the thermodynamic limit:
\begin{equation}
0<\Delta_{N\rightarrow\infty}=1-\alpha <1\;.
\end{equation}
Using~(\ref{distance}), this result immediately implies fast scrambling for this model, since the relaxation time is given by
\begin{equation}
t_{\textrm{relaxation}}\sim \log N\;.
\end{equation}
Going beyond this concrete result, it is tempting to speculate that the non-vanishing gap of this democratic walk is dual to the non-vanishing gap characteristic of the near horizon hyperbolic geometry of thermal event horizons, or their discrete version, the so-called expander graphs \cite{avi1}, highlighted in \cite{us1,us3} in the context of the fast scrambling conjecture. We hope to come back to this interesting connection in future work.

Another interesting hierarchy of values is the following:
\begin{equation}
\eta=\frac{1}{N\log N}\;,
\end{equation}
and therefore
\begin{equation}
\alpha=1-(N-1)\,\eta=1-\frac{N-1}{N\log N}\;.
\end{equation}
Again, using the result~(\ref{eigen}), in the thermodynamic limit the second eigenvalue $\lambda\rightarrow\alpha$, since $\eta\rightarrow 0$ in this limit. But in this case, the second eigenvalue is not bounded away from $1$, since  it is given by
\begin{equation}\label{ultra}
\lambda \rightarrow\alpha\rightarrow1-\frac{1}{\log N}.
\end{equation}
Although in this case the relaxation time of the democratic walk is rigorously of order $(\log N)^{2}$, the time scale $\log N$ is still meaningful, as we will comment in more detail later. Indeed, within this heirarchy we will be able to qualitatively obtain the results found in \cite{stanford2014,sahakian2,joansimon}, and also the estimate done in section~(\ref{sec5}), while still giving a coherent picture of information spreading at the global level. Given relation~(\ref{ultra}), this family of democratic walks resembles the ultrametric diffusion processes presented in \cite{us4}.

We conclude that within the window $\mathcal{O}(1/N)<\eta<\mathcal{O}(1/N\log N)$, the democratic walk is non trivial and furthermore shows fast scrambling features.

\subsection{Randomized democratic walks}

In the previous section we used a simplified model in which the entries of the transition matrix were fixed, see~(\ref{rel1}) and~(\ref{rel2}). The objective was to find a toy model in which everything was explicitly solvable. But this toy model would not be of too much help if it was sensitive to small perturbations of the transition matrix entries. Indeed, in more realistic situations each entry of the transition matrix would be somewhat different from the others, each one corresponding to some transition/correlator between different states/operators and also coupling dependent. We thus can ask if the previous results are robust, and if they remain generically valid for any kernel with a prescribed hierarchical structure. 

To answer this question we add a random matrix $R$, taken from the GOE ensemble\footnote{See \cite{tao} for a nice and complete treatment of random  matrices, including the GOE ensemble.}, to the previous transition matrix $M$. Being taken from the GOE ensemble means that the added $R$ is a symmetric matrix with real, independent and identically distributed random coefficients, characterized by $[R_{ij}]=0$ and $[R_{ij}^{2}]=\sigma^{2}=\eta^{2}$, where the previous brackets define the averages of the random matrix entries, and indeed define the GOE ensemble itself. Notice that the GOE ensemble is the right ensemble to choose, since the transition matrix $M$ is symmetric and has real coefficients. Notice also that setting $\sigma = \eta$ ensures the prescribed hierarchy. The result is that the full transition matrix
\begin{equation}
\bar{M}=M+R\;,
\end{equation}
satisifies $[\bar{M}]=M$, but it has been randomly kicked out from the mean by perturbations with a typical size of $\mathcal{O}(\eta)$. Therefore, the random matrix $\bar{M}$  has the hierarchical structure defined by relations~(\ref{rel1}) and~(\ref{rel2}), but otherwise it has random entries. The objective is to show that the eigenvalues are stable/robust to such perturbations, and that the hierarchy window for fast scrambling walks found in the previous section remains intact.

To study the spectrum of $\bar{M}$ we use the basic theory of random matrices together with the so-called Weyl inequalities, which are inequalities constraining the eigenvalues of the sum of two matrices with the use of the eigenvalues of each of the separate matrices\footnote{For an extensive and self-contained treatment of random matrices and these inequalities we refer to \cite{tao}.}. In particular, if $\lambda_{i}(A+B)$, $\lambda_{i}(A)$ and $\lambda_{i}(B)$ are the eigenvalues of the matrices $A+B$, $A$ and $B$ respectively, with sizes $N\times N$, ordered as $\lambda_{1}\geq\lambda_{2}\geq\cdots \geq\lambda_{N}$, then we have the so-called Weyl inequality:
\begin{equation}
\lambda_{i+j-1}(A+B)\leq\lambda_{i}(A)+\lambda_{j}(B)\;,
\end{equation}
valid whenever $i+j-1\leq N$, and the dual Weyl inequality:
\begin{equation}
\lambda_{i+j-N}(A+B)\geq\lambda_{i}(A)+\lambda_{j}(B)\;,
\end{equation}
valid whenever $i+j-N\leq N$. Together they imply the following useful inequality:
\begin{equation}
\lambda_{i}(A)+\lambda_{N}(B)\leq\lambda_{i}(A+B)\leq\lambda_{i}(A)+\lambda_{1}(B)\;,
\end{equation}
valid for any $i$. Related to these inequalities we have the so-called `eigenvalue stability inequality', which reads:
\begin{equation}
\vert\lambda_{i}(A+B)-\lambda_{i}(A)\vert \leq \vert B\vert_{\textrm{op}}\;,
\end{equation}
where $\vert B\vert_{\textrm{op}}$ is the operator norm of the matrix, defined as $\vert B\vert_{\textrm{op}}:=\textrm{sup}\vert B v\vert$, where the symbol `$\textrm{sup}$' means we should consider all vectors $v$ with unit norm and take the maximum $\vert B v\vert$.

For our case $A=M$, defined by~(\ref{rel1}) and~(\ref{rel2}), with eigenvalues given by $1$ and $\alpha-\eta$ as derived in the previous section, and $B=R$, a random matrix taken from the GOE ensemble. One of the most basic/fundamental results in random matrix theory is Wigner's semicircle law, see \cite{tao,avi1},  which constrains the eigenvalues of $R$ in the large $N$ limit to live in a compact domain:
\begin{equation}
\lambda^{2}\leq 4 N \sigma^{2}\;,
\end{equation}
where we remind that $[R_{ij}^{2}]=\sigma^{2}$ is the mean squared deviation of each matrix entry. For $\sigma^{2}=\eta^{2}$ we get $-2\,\sqrt{N}\,\eta\leq\lambda_{i}(R)\leq 2\,\sqrt{N}\,\eta$.
Another important and related result in random matrix theory is the computation of the expected operator norm, which is given by $[\vert R\vert_{\textrm{op}}]=2\, \sqrt{N}\, \sigma =2\,\sqrt{N}\,\eta$.

Using Weyl inequalities or the eigenvalue stability inequality, we arrive at:
\begin{equation}
\vert\lambda_{i}(\bar{M})-\lambda_{i}(M)\vert\leq 2\,\sqrt{N}\,\eta\;,
\end{equation}
which, together with the results of the previous section, implies that the hierarchy window with fast scrambling behavior remains unchanged. In particular we have the specific class of democratic walks with non-vanishing gap for $\eta\sim \mathcal{O}(1/N)$, akin to expander graph diffusion, and the specific class of democratic walks with $\eta\sim \mathcal{O}(1/N\log N)$, akin to ultrametric behavior.

Finally we remark that the random matrix perturbation does not spoil the specific way in which the initial probability distribution evolves to stationarity. In particular,
\begin{equation}\label{solution2}
p^{n}=\bar{M}^{n} p^{0}=\begin{pmatrix}
\bar{\alpha}_{n} \\ \bar{\eta}_{n} \\ \bar{\eta}_{n} \\ \vdots \\ \bar{\eta}_{n}
\end{pmatrix}
\;,
\end{equation}
where now $\bar{\alpha}$ and $\bar{\eta}$ are random variables with mean values $\alpha$ and $\eta$ respectively, and with squared deviations given by $\sigma^{2}=\eta^{2}$.

We conclude that all probability entries, except for the one corresponding to the perturbation, are relaxed after one step, signalling the simple structure of information leaking in democratic models. We only need to wait for the initially non-zero entry to decay to a prespecified value.

Notice also that the results would not change if the diagonal entries were different between each other, but still all of them of $\mathcal{O}(\alpha)$. This would correspond to the more generic situation in which each degree of freedom has a different decay rate. Still, this variation would not change the main conclusions.

\section{Scrambling as stochastic evolution of Information fields}\label{sec3}

A recurring problem in quantum thermalization and information mixing/scrambling is to find computational frameworks allowing the description of information flows in the process of unitary evolution. One appealing approach is to identify some locally conserved charge, such as energy and momentum in translationally invariant theories, and use an effective random walk or diffusion process to describe its location as time evolves. This approach has two important disadvantages, of direct relevance in the context of black hole physics. The first one is that not all models have such simple conserved charges, so it is not clear whether random walks represent the evolution of some physical quantity associated to the theory. The second one is that, whenever such a conserved charge exists, it is not clear how to relate the relaxation time of the effective random walk with more subtle information theoretic time-scales, such as the scrambling time \cite{hayden, susskind}.

In this section we connect stochastic processes, information flows and unitary quantum dynamics in a direct way. We will show how to construct various notions of information density distributions, or {\it information fields} $\mathcal{I}$, for any given many-body quantum system.  These information fields will be associated to a subsystem $\textbf{A}$ and a quantum state $\rho$, so we denote them by $\mathcal{I}_{\textbf{A}}^{n} (\rho)$, where $n=1, \cdots , N$ runs over the different degrees of freedom of the system. We will show that microscopic unitary evolution induces an unambigous stochastic evolution of the information field, and that the final state of a non-equilibrium process is always the uniform distribution, whatever the initial state. As a byproduct, we will show that the construction transparently connects spreading of information with growth of entanglement entropy.

The approach then avoids the previous problems. The stochastic evolution of the information field  $\mathcal{I}_{\textbf{A}} (\rho)$ provides a possible and rigorous operational definition of `information spreading', a type of quantum transport,  in many-body quantum systems. In the context of non-equilibrium unitary processes, it gives a rigorous notion of information scrambling \cite{susskind}. Besides, the information field always exists, even when locally conserved charges do not. It is tempting to speculate that this approach could potentially turn into a generic description of chaos in many body systems, whether classical or quantum, as the classical mixing properties of the stochastic evolution of the information density.

Finally, notice that in a non-equilibrium process, given the stochasticity of the evolution of the information field, one can take a Markovian approximation, which is expected to hold to some extent given the thermal nature of the process. Although the proper justification of the Markovian approximation will be left unclear\footnote{Notice that this approximation is also unproven for more conventional densities such as energy densities or charged currents.}, we will see that it will allow us to obtain specific laws for the evolution of entanglement entropy. In particular it will allow us to obtain qualitatively the results presented in \cite{stanford2014,sahakian2,joansimon}, together with the arguments developed in section~(\ref{sec5}).

To construct the information field we will use the framework and ideas developed in \cite{uscod}. In the first part of this section we review that construction.


\subsection{Codification Volumes}\label{cod}

In this section we review the novel approach to study quantum thermalization developed in \cite{uscod}, which serves as the basis for the construction of the information field. In this approach, the task is not to study the relaxation properties of a given subsystem, but to ask for the `location' in the global pure state of the information lost by the subsystem. To use a very simple but illustrative example taken from \cite{uscod}, consider the following pure state of a system of $N$ spins:
\begin{equation}
\vert\psi\rangle=\vert\uparrow\rangle_{1} \otimes \vert\varphi\rangle_{2, \cdots , N}\;,
\end{equation}
where the numbers $1,\cdots, N$ label the spins.  The specific $\vert\varphi\rangle_{2,\cdots, N}$ will not be important for the argument, just the fact that there is a large number of spins. Imagine we are interested in the first spin, with reduced state\footnote{The reduced state is obtained by tracing out the complement in the usual manner, see \cite{preskillnotes} for a nice presentation.} denoted by $\rho_{1}$, and that unitary evolution produces the following pure state:
\begin{equation}\label{twostateN}
\vert\psi\rangle=\frac{1}{\sqrt{2}}(\vert\uparrow\downarrow\rangle_{1,2}-\vert\downarrow\uparrow\rangle_{1,2}) \otimes \vert\varphi\rangle_{3,\cdots, N}\;.
\end{equation}
If we were to measure the properties of $\rho_{1}$, we would conclude that the state is thermal with $\beta=1/T=0$, no matter if we use local operators (the Pauli matrices) or entanglement entropy. At the same time, it is clear that information about spin $1$ has yet not been \emph{thermalized/scrambled} at all, considering that the spin $1$ is only correlated with spin $2$, and not with any spin in the set $3, \cdots ,N$. We conclude that it is important to ask where the information about a subsystem has been located on the full quantum state in order to conclude that the state is fully thermalized/scrambled.

We are led to the generic question of whether the information associated to a subsystem $A$ can be ``localized'' in a different subsystem $B$, and if there is a definite procedure to find this $B$. 
Even if subsystem $A$ is entangled with its complement, one might expect that in many cases (as in the previous example) most of its information is distributed only within a certain subsystem $B$. The natural question is then for the minimum size $B$ where all the information of $A$ can be found. In the previous example the answer would be the second spin. With the objective of formulating this question quantitatively, in \cite{uscod} the quantity $\Omega_{\textbf{A}}(\rho)$, called the \emph{Codification Volume} of an operator algebra $\textbf{A}$, was introduced \footnote{To define properly the codification volume one needs to introduce a precision $\epsilon$. In this article we will not need those technicalities, and refer to \cite{uscod} for a more rigorous treatment.}.

Before reviewing the definiton of $\Omega_{\textbf{A}}(\rho)$, notice that if we want to measure the total correlation between two operator algebras $\textbf{A}$ and $\textbf{B}$ (equivalently two subsystems $A$ and $B$ in the case of factorizable Hilbert spaces), we need to find the distance of the associated reduced density matrix $\rho_{\textbf{A}\textbf{B}}$ to a factorized state $\rho_{\textbf{A}}\otimes
\rho_{\textbf{B}}$ \footnote{We want to remark that the reduced density matrix of an operator algebra is perfectly defined for factorizable Hilbert spaces, being just the density matrix of the associated subsystem. But indeed it works nicely for more generic cases in which the operator algebra does not define a partition of the Hilbert space \cite{amilcar1,amilcar2}. This may indeed be important for applications to black holes, since dual gauge theories do not posse factorizable Hilbert spaces.}. For this task we use the quantum relative entropy $S(\rho\Vert \sigma)$, defined by
\begin{equation}\label{relativedefinition}
S(\rho\Vert \sigma)=\textrm{Tr}\rho (\log\rho-\log\sigma)\;,
\end{equation}
because it is a higher bound of all other possibilities \cite{eisert,cirac}. In the case we are interested $A$ and $B$ are disjoint subsystems, and the relative entropy coincides with the Mutual Information (MI) $I(\textbf{A},\textbf{B})$:
\begin{equation}\label{relativeI}
S(\rho_{\textbf{A}\textbf{B}}\Vert \rho_{\textbf{A}}\otimes 
\rho_{\textbf{B}})= S_{\textrm{E}}(\rho_{\textbf{A}})+S_{\textrm{E}}(\rho_{\textbf{B}})-S_{\textrm{E}}(\rho_{\textbf{A}\textbf{B}})= I(\textbf{A},\textbf{B}).
\end{equation}
The MI gives the total amount of correlations between two systems \cite{groisman2005,cirac}. It is a measure of how much we can learn about $\textbf{A}$ by studying $\textbf{B}$ and vice versa.

Given these linguistic preliminaries, the Codification Volume of an operator algebra $\Omega_{\textbf{A}}(\rho)$ was defined by
\begin{equation}\label{omega}
\Omega_{\textbf{A}}(\rho)\equiv \textrm{Min} ~~ S_{\textbf{B}}\;,
\end{equation}
where the right hand side of the equation is the number of degrees of freedom of $\textbf{B}$, and where the minimum, running over all subsystems $B$ with associated $S_{\textbf{B}}$, is taken such that the following relation holds:
\begin{equation}\label{eq:cond}
I(\textbf{A},\bar{\textbf{A}})-I(\textbf{A},\textbf{B})\simeq 0 \;,
\end{equation}
where $\bar{\textbf{A}}$ is the algebra complementary to $\textbf{A}$. The symbol~`$\simeq$'~is irrelevant here. It originates from the needed precision $\epsilon$ to define rigorously the Codification Volume. But this $\epsilon$ obscures the physics involved in equation~(\ref{eq:cond}), so we will omit it in what follows and refer to \cite{uscod} for the proper treatment.

The intuitive interpretation is that when condition~(\ref{eq:cond}) holds, all the information shared by $\textbf{A}$ might be found \emph{just} by looking at $\textbf{B}$.

Notice that if the global state is pure, equation~(\ref{eq:cond}) can be rewritten as:
\begin{equation}
I(\textbf{A},\textbf{B})\simeq 2 S_{\textrm{E}}(\textbf{A})\;,
\end{equation}
The right hand side is the full amount of information that has been shared by $\textbf{A}$, equal to $I(\textbf{A},\bar{\textbf{A}})= 2 S_{\textrm{E}}(\textbf{A})$. The left hand side means that all this information can be found by looking at correlations between $\textbf{A}$ and $\textbf{B}$. In the previous example~(\ref{twostateN}) one can check that indeed $I(\textbf{1},\bar{\textbf{1}})-I(\textbf{1},\textbf{2})=0$, showing that all information can be found just by looking at the second spin.

One of the reasons for our interest in the previous quantity is that $\Omega_{\textbf{A}}(\rho)$ is extensive for random states \cite{uscod}. In a non-equilibrium unitary process, if the initial state is a factorized state:
\begin{equation}
\vert\Psi\rangle_{1,\cdots , N}=\vert\psi\rangle_{1}\otimes\vert\psi\rangle_{2}\otimes\cdots\otimes\vert\psi\rangle_{N}\;,
\end{equation}
$\Omega_{\textbf{1}}(\rho)$ increases from $0$ to $\mathcal{O}(N)$ in the course of thermalization \footnote{Here we want to remark that this is true for any prespecified, or time independent, basis of the Hilbert space, not equal to the eigenstate basis of course. But this is as generic as one can get, since there is no notion of information flow without a prespecified basis. In other words, if the state of the system is $\rho(t)$ and one chooses a time dependent basis with $\rho(t)$ being one of the basis vectors, there is no information flow whatsoever between the different basis vectors. This is trivial and uninteresting, and it ceases to be true once we choose a time independent basis.}. The growth can be measured by the defining relation
\begin{equation}\label{evolcod}
I(\textbf{A},\textbf{B}(t))\simeq I(\textbf{A},\bar{\textbf{A}})\;,
\end{equation}
which in the case of a unitarily evolving system is equivalent to
\begin{equation}\label{unitarity}
I(\textbf{A},\textbf{B}(t))\simeq 2 S_{\textrm{E}}^{\textbf{A}}\;.
\end{equation}
This was studied in \cite{uscod} for the case of a one dimensional spin chain, showing explicitly how information flows through the system. The growth of $\Omega_{\textbf{1}}(\rho)$ allows a rigorous study of information spreading in non-equilibrium processes, even in cases where there are no conserved charges arising from symmetries.

In the next subsection we show how to use the previous approach to construct certain `information densities' over the set of degrees of freedom.

\subsection{Stochastic evolution of information fields}\label{stoinfo}

Assume we are able to solve equation~(\ref{evolcod}), and find the minimum $\textbf{B}$'s with whom $\textbf{A}$ is maximally correlated \footnote{Notice that in some cases, there is only one minimum subsystem maximally correlated with a given subsystem $\textbf{A}$, like in the example given in the previous section~(\ref{twostateN}). But this is not the general rule within quantum mechanics. An example with direct relevance for our construction is the case of a random state \cite{uscod}. The Codification Volume for the random state turns out to be equal to $N/2$. But indeed every subsystem with $N/2$ degrees of freedom is maximally correlated with $\textbf{A}$. It is clear why, when defining the information density, we will divide the lost information over all the possible minimum $\textbf{B}$ maximmally correlated with $\textbf{A}$, i.e satisfying equation~(\ref{evolcod}). In the case of a random state this implies we should democratically distribute the lost information over the $b=N-a\geqslant\frac{N}{2}$ degrees of freedom not belonging to $\textbf{A}$.}. Knowing the set of $\textbf{B}$'s maximmally correlated with $\textbf{A}$ allows the definition of a density of Mutual Information associated to the subsystem $\textbf{A}$, which we will term an information field $\mathcal{I}_{\textbf{A}}(\rho)$. This information field provides a measure of how the information about $A$ is distributed over the different degrees of freedom of the system in the global quantum state $\rho$.

If the number of degrees of freedom in $\textbf{A}$ is `$a$', the set of $\textbf{B}$'s have altogether `$b$' degrees of freedom, and $\mathcal{I}_{\textbf{A}}(\rho)$ is organized such that its first `$a$' entries correspond to the degrees of freedom in  $\textbf{A}$, its second `$b$' entries correspond to the ones in all $\textbf{B}$'s satisfying~(\ref{evolcod}), and the rest of the entries for the left over degrees of freedom, we define the information density associated to $\textbf{A}$ as \footnote{It is obviously unimportant how we organize the degrees of freedom in the information field. We just do it this way for notational and visual convenience, so that its properties become more transparent.}
\begin{equation}\label{density}
\mathcal{I}_{\textbf{A}}(\rho)=\frac{1}{S_{\textrm{E}_{\textrm{s}}}^{\textbf{A}}}\left(\begin{array}{rcl}  &\frac{S_{\textrm{E}_{\textrm{s}}}^{\textbf{A}}-S_{\textrm{E}}^{\textbf{A}}}{a}+\frac{I(\textbf{A},\textbf{B})}{2(b+a)} &\\ & \vdots & \\  &\frac{S_{\textrm{E}_{\textrm{s}}}^{\textbf{A}}-S_{\textrm{E}}^{\textbf{A}}}{a}+\frac{I(\textbf{A},\textbf{B})}{2(b+a)} &\\ &\frac{I(\textbf{A},\textbf{B})}{2(b+a)}&\\ &\vdots & \\ &\frac{I(\textbf{A},\textbf{B})}{2(b+a)} & \\ & 0 & \\ &\vdots & \\ & 0 &
\end{array}\right) \;,
\end{equation}
where $S_{\textrm{E}_{\textrm{s}}}^{\textbf{A}}$ is the entanglement entropy at the stationary regime, associated to the reduced density matrix of the subsystem $\textbf{A}$ in the stationary state, and $S_{\textrm{E}}^{\textbf{A}}$ is the entanglement entropy of $\textbf{A}$ in the actual state $\rho$. Usually, but not always, the unitary evolution drives the global pure state to a stationary regime in which the reduced density matrix is approximately thermal at inverse temperature $\beta$, see \cite{usrandom} for a recent treatment. In this case the entanglement entropy at stationarity $S_{\textrm{E}_{\textrm{s}}}^{\textbf{A}}$ is just the thermal entropy of the subsystem $\textbf{A}$ \footnote{The defintion is specifically thought for non-equilibrium processes with stationary regimes, the ones we are interested in this article. But otherwise there might be related definitions of conceptually equivalent Information fields more suited for other situations. It would be interesting to explore these issues further.}. 

With the definiton at hand, we now want to discuss various aspects associated to $\mathcal{I}_{\textbf{A}}(\rho)$. The first is that given the state of the system $\rho (t)$ at any time $t$, the information field $\mathcal{I}_{\textbf{A}}(\rho (t))$ can be found unambiguously. Therefore, the evolution of $\rho (t)$ generates a well defined evolution for the information field, given by $\mathcal{I}_{\textbf{A}}(\rho (t))$. Secondly, if the initial state $\rho (t_{\textrm{in}})$ is pure and the microscopic evolution of $\rho$ is unitary, the corresponding evolution $\mathcal{I}_{\textbf{A}}(\rho (t))$ is stochastic, in the sense that the normalization of the information field is preserved through unitary evolution\footnote{Not only the normalization is conserved, but all the entries are non-negative, so the information field is matematically equivalent to a probability distribution. We thank Jos\'e L.F. Barb\'on for signalling the importance of this fact when constructing distributions of this type.
 } . This is rooted in relation~(\ref{unitarity}). Mathematically we have:
\begin{equation}\label{uni}
\rho (t) = U(t-t_{\textrm{in}})\vert\psi (t_{\textrm{in}})\rangle\langle\psi (t_{\textrm{in}})\vert U^{\dag}(t-t_{\textrm{in}})\Rightarrow I(\textbf{A},\textbf{B}(t))\simeq 2 S_{\textrm{E}}^{\textbf{A}}\Rightarrow \sum_{n=1}^{N}\mathcal{I}_{\textbf{A}}^{n}(\rho (t))=1\;.
\end{equation}
As in the previous section, the symbol~`$\simeq$'~is irrelevant here. It again originates from the needed precision $\epsilon$ to define rigorously the Codification Volume, as done in \cite{uscod}. It can be properly included in the definition of $\mathcal{I}_{\textbf{A}}(\rho (t))$. But this technical detail does not seem to bring any difference in practice. It just makes the definition and properties of the information density $\mathcal{I}_{\textbf{A}}(\rho (t))$ more obscure and difficult to understand, so we will not consider it in what follows.

The intuition behind the vector~(\ref{density}) is the following. In a non-equilibrium process, the final amount of information which will be shared by subsystem $\textbf{A}$ with the environment is simply given by the entanglement entropy at stationarity $S_{\textrm{E}_{\textrm{s}}}^{\textbf{A}}$. This quantifies the information lost by $\textbf{A}$ due to interactions with the environment. So a measure of the information about $\textbf{A}$ stored in the degrees of freedom of $\textbf{A}$ itself is given by the total $S_{\textrm{E}_{\textrm{s}}}^{\textbf{A}}$ minus the amount that has been lost $S_{\textrm{E}}^{\textbf{A}}(t)$, plus the amount that has been shared with all minimal $\textbf{B}$'s, which is 
$\frac{I(\textbf{A},\textbf{B})}{2(b+a)}$. On the other hand, the degrees of freedom belonging to the set of $\textbf{B}$'s have gained an amount of information $\frac{I(\textbf{A},\textbf{B})}{2(b+a)}$ about $\textbf{A}$.

Having shown that the evolution of $\mathcal{I}_{\textbf{A}}(\rho (t))$ is stochastic given a unitary evolution of the microscopic quantum state, we now want to show that in non-equilibrium thermal processes the final state is the stationary distribution $\mathcal{I}_{\textbf{A}}^{n}(\rho (t\gtrsim t_{\textrm{relax}}))\simeq 1/N$. Indeed, in these processes, as time evolves the entanglement entropy approaches the stationary regime $S_{\textrm{E}}^{\textbf{A}}\rightarrow S_{\textrm{E}_{\textrm{s}}}^{\textbf{A}}$, by definition of $S_{\textrm{E}_{\textrm{s}}}^{\textbf{A}}$. On the other hand, from the results in \cite{uscod} we conclude that, in the stationary regime of such a thermal process, all minimum $\textbf{B}$ with size $N/2$ satisfy relation~(\ref{evolcod}). Therefore we have $b=N-a$ at stationarity, and $\mathcal{I}_{\textbf{A}}^{n}(\rho (t\gtrsim t_{\textrm{relax}}))\simeq 1/N$. More properly, the precise mathematical statement is that the Information field in a random state, i.e. the \emph{typical} information field, is given by $\mathcal{I}_{\textbf{A}}^{n}(\rho_{\textrm{random}})=1/N$.

Given this construction and properties of $\mathcal{I}$, if we have a many-body quantum system with $N$ degrees of freedom, and the initial quantum state is factorized\footnote{It is simple to repeat the same argument in the case of a quantum quench, in which the initial quantum state is a vacuum state of certain local theory. The only difference would be that the initial information field would have more than one non-zero entry. It seems that this construction point to the factorized state as the state providing the biggest scrambling time, and so the type of state which needs to be considered for stablishing bounds.}
\begin{equation}
\vert \Psi_{t=0}\rangle =\vert\psi_{1}\rangle\otimes\vert\psi_{2}\rangle\otimes\cdots\otimes\vert\psi_{N}\rangle\;,
\end{equation} 
the initial state of the information field associated to subsystem $\textbf{1}$ can be easily found through the definition, and it is given by:
\begin{equation}
\mathcal{I}_{\textbf{1}}(\rho (t=0))=\left(\begin{array}{rcl}  &1 &\\&0 &\\ &\vdots &  \\ & 0 &
\end{array}\right)\;.
\end{equation} 
Through unitary evolution, the properties of the exactly evolved quantum state are going to be similar to those of a typical state \footnote{See \cite{usrandom} for a recent discussion in the QFT context.}
\begin{equation}\label{randomuni}
\vert \Psi_{t=0}\rangle\longrightarrow\vert \Psi_{\textrm{random}}\rangle\:,
\end{equation}
so that
\begin{equation}\label{sto}
\mathcal{I}_{\textbf{1}}(\rho (t=0))=\left(\begin{array}{rcl}  &1 &\\&0 &\\ &\vdots &  \\ & 0 &
\end{array}\right)\longrightarrow \left(\begin{array}{rcl}  &1/N &\\&1/N &\\ &\vdots &  \\ & 1/N &
\end{array}\right)=\mathcal{I}_{\textbf{1}}^{\textrm{typical}}\;.
\end{equation} 
The stochastic evolution of $\mathcal{I}_{\textbf{A}}(\rho (t))$ towards the uniform distribution provides a rigorous, but still operative definition of information mixing/scrambling in many body quantum systems, and it is one of the main results of the present work. Characteristic time scales of this stochastic process provide notions of the scrambling time in the quantum system \cite{hayden, susskind}, as we describe further below\footnote{Notice that the ballistic spreading of information found generically in local systems, such as spin models or CFT's, does not cause any contradiction, since the Markovian evolution does not imply diffusion laws. We can always include phenomenological drift velocities in the Markov process, and produce ballistic behavior. Derivation of these drift terms in this information theoretic perspective is an interesting path to further explore.}. Notice that in this formulation, growth of entanglement entropy and spreading of information through the system are highly interrelated, since the entries of the information field contain the entanglement entropies themselves. We will show in more detail how this connection works in the next section.

We want to remark that this framework is valid even if there is no local conserved charge related to some symmetry of the system. The purity of the global state ensures that the increase of entanglement entropy $S_{\textrm{E}}^{\textbf{A}}(t)$ of the subsystem under study, from the initial value to the final stationary value $S_{\textrm{E}_{\textrm{s}}}^{\textbf{A}}$, is mirrored by an equal increase of the entanglement entropy of the complementary subsystem $S_{\textrm{E}}^{\bar{\textbf{A}}}(t)=S_{\textrm{E}}^{\textbf{A}}(t)$. But $S_{\textrm{E}}^{\bar{\textbf{A}}}(t)$ is not neccessarily democratically spread over $\bar{A}$, and effectively behaves as a conserved charge spreading through $\bar{A}$. The evolution and localization of entanglement spreading is carefully followed by  $\Omega_{\textbf{A}}(\rho)$, or more transparently by $\mathcal{I}_{\textbf{A}}(\rho)$.

To resume, we have defined an unambiguous information density or information field $\mathcal{I}_{\textbf{A}}(\rho (t))$ associated to any given $\textbf{A}$. We have shown that microscopic unitary evolution translates into stochastic evolution of this density. This is remarkable, since the stochastic process generically does not conserve the Shannon entropy associated to the information density, while unitarity conserves the Von Neumann entropy of the global quantum state. Besides, for unitary processes with random stationary regimes (non-equilibrium thermal processes), the unitary time evolution drives the information field to the usual stationary distribution, given by $\mathcal{I}_{\textbf{A}}^{n}(\rho (t\gtrsim t_{\textrm{relax}}))\simeq 1/N$. In this framework, quantum chaotic phenomena can be studied by the mixing properties of the classical stochastic evolution of the information field.

Finally, although it is not neccessarily true that the evolution of $\mathcal{I}_{\textbf{A}}(\rho (t))$ for $t>t_{0}$ just depends on $\mathcal{I}_{\textbf{A}}(\rho (t_{0}))$, we expect the Markovian approximation to hold to some extent in some regimes, at least for non-equilibrium unitary proccesses, akin to thermal dynamics \footnote{To give a simple example, in \cite{uscod} the growth of the codification volume was seen to follow a smooth linear law, with a probable effective description in terms of a differential equation local in time. In previous literature, effective and local in time equations have been proposed for the growth of entanglement entropy, see \cite{esperanza2010}. Besides, notice that the same problem appears in other thermodynamical equations, such as diffusion of conserved charges, in which the same Markovian approximation is needed to arrive to the final classical diffusion equations.}. We thus conclude that this framework justifies the use of Markov processes to study information spreading in many-body quantum systems. To the author's knowledge, it gives a new connection between quantum and classical equations of motion, a feature which deserves more development. Besides, as we will see in the next section, given a specific kernel structure for the random walk, the evolution of this density provides specific laws for the evolution of entanglement entropies in the system. This will provide a direct connection between spreading of correlations and growth of entanglement.

\section{Democratic information flows for democratic physical systems.}\label{sec4}

In this section we want to use the previous formalism in the context of democratic systems. We also want to combine it with the specific democratic walk model we described in section~(\ref{sec2}). With democratic systems we mean systems with $N$ degrees of freedom, characterized by bounded or unbounded operators, for which the Hamiltonian contains interaction terms connecting every degree of freedom with every other.

As commented before, when combining the two frameworks, we need to take a Markovian approximation. With this in mind, in the first part of this section we consider the generic implications of democratic interactions in the stochastic evolution of the information field, without any Markovian assumption.

In the second part we analyze the implications of assuming Markovian evolution and using the democratic walks described in section~(\ref{sec2}) as kernels of the information field. This last subsection is therefore speculative, since we cannot prove that the stochastic evolution of the information field $\mathcal{I}$ is indeed given by one specific democratic walk. At any rate we will show that for one specific hierarchy, the democratic walk qualitatively reproduces the results presented in \cite{stanford2014,sahakian2,joansimon}, together with the bulk estimate done in section~(\ref{sec5}), while still transparently showing the information spreading properties of democratic systems.

Lastly we also comment on the thermodynamic limit of the process, which sheds new light on the fate of information in black hole physics.

\subsection{Generic features} 

Without assuming any approximation, we can infer the generic structure of information leaking in democratic models by the following argument. The key aspect to notice is that in a system with democratic interactions, the connected correlations are neccessarily democratic, since they are functions of the democratic couplings themlseves. For example, if we begin with a factorized state:
\begin{equation}
\vert \Psi_{t=0}\rangle =\vert\psi_{1}\rangle\otimes\vert\psi_{2}\rangle\otimes\cdots\otimes\vert\psi_{N}\rangle\;,
\end{equation} 
democratic interactions imply that at time $t$ we must have the following chain of  relations:
\begin{eqnarray}\label{democracy}
&I(\textbf{1},\textbf{2})(t)\simeq  I(\textbf{1},\textbf{3})(t)\simeq \cdots\simeq I(\textbf{1},\textbf{N})(t) \;;&\nonumber \\
&I(\textbf{1},\textbf{2}\textbf{3})(t)\simeq  I(\textbf{1},\textbf{2}\textbf{4})(t)\simeq \cdots\simeq I(\textbf{1},(\textbf{N-1})\textbf{N})(t)\;;&\nonumber\\
&\cdots &\;,
\end{eqnarray}
where equality holds for a model with exact democratic interactions. The previous chain of relations means that the mutual information, or otherwise the total connected correlations between the first degree of freedom and a block of `$m$' degrees of freedom just depends on the size `$m$', and not on the specific degrees of freedom chosen. This is because the evolution of mutual informations/connected correlations is finally determined by the couplings between the different degrees of freedom, and these are democratic by assumption. If one mutual information in~(\ref{democracy}) is significantly bigger than the others, and evolves in a qualitatively different way, this must be rooted in a hierarchy of couplings in the system, contradicting the assumption of democratic interactions. So even without knowing the exact relation between the mutual information growth functions and the interaction couplings, relation~(\ref{democracy}) holds \footnote{This is exactly analogous to the statement that in local euclidean theories with rotational invariance, information flows in a rotationally invariant way approximately, and it is exactly rotationally invariant if the initial state is rotationally invariant too.}.

In turn, relation~(\ref{democracy}) implies that if we find a subset $\textbf{B}$ of degrees of freedom with whom $\textbf{1}$ is maximmally entangled:
\begin{equation}
I(\textbf{1},\textbf{B})\simeq 2 S_{\textrm{E}}^{\textbf{1}}\;,
\end{equation}
then any other subsystem as big as $\textbf{B}$ will also be maximally entangled with $\textbf{1}$. Joining this aspect, coming from the assumed democracy of interactions of the quantum model, with the fact that the information has to be \emph{somewhere}, due to unitarity of the evolution, democracy of interactions implies that the information field  evolves as:
\begin{equation}\label{NoMarkov}
\mathcal{I}_{\textbf{1}}(\rho (t))=\begin{pmatrix}
\alpha_{t} \\ \eta_{t} \\ \eta_{t} \\ \vdots \\ \eta_{t}
\end{pmatrix}\;,
\end{equation}
Notice that the entry corresponding to the first degree of freedom is not constrained by~(\ref{democracy}) to be of the same order of magnitude as the others. This aspect is crucial, as we will explain later on. So, even without assuming the Markovian approximation, relation~(\ref{NoMarkov}) holds and expresses the simple structure of information leaking in democratic models. We conclude:
\begin{itemize}
\item Scrambling in democratic systems amounts to `local' relaxation of the initial perturbation, the information spreading process being instantaneous. 
\end{itemize} 
We just need to agree about `how much' information we want to obtain from $\textbf{1}$, and then compute the relaxation time needed for this specific leaking. There are two natural possibilities. The first is to ask for stabilization of the stochastic process. This convention reads:
\begin{equation}\label{NoMarkovcond}
R=\frac{ \eta_{t}}{\alpha_{t}}\sim \mathcal{O}(1)\;.
\end{equation}
The second is to ask for extensivity of entanglement entropy. Notice that from the definition of the information field~(\ref{density}), for a democratic model we have:
\begin{equation}
\frac{S_{\textrm{E}}^{\textbf{1}}(t)}{N S^{\textbf{1}}_{\textrm{E}_{\textrm{s}}}}=\eta (t)\;,
\end{equation}
where $S_{\textrm{E}}^{\textbf{1}}(t)$ is the evolution of entanglement entropy of the first degree of freedom, and $S^{\textbf{1}}_{\textrm{E}_{\textrm{s}}}$ is the entanglement entropy at stationarity of the first degree of freedom. We see that in this formalism, spreading of information and evolution of entanglement entropy are tightly related. Asking now for extensivity amounts to ask for:
\begin{equation}\label{extensivity}
\eta(t)\sim \mathcal{O}(1/N)\;.
\end{equation}
For local systems the first convention is the right one, since what it is non trivial is the quantum transport of information. For democratic models it is not completely clear. Regarding the physics of the fast scrambling conjecture, we will argue below that the most appropriate is the second one.

Lastly, from the perspective we are looking at the problem it is natural to assume that the information flow do not strongly depend of its initial location. Assuming this is the case, from~(\ref{NoMarkov}) and~(\ref{democracy}) we can get another generic property of democratic systems, not depending on the specific functional form of $\eta(t)$. By repeating the same argumentation for a subsystem $\textbf{A}$ of `$a$' degrees of freedom we arrive to:
\begin{equation}
\frac{S_{\textrm{E}}^{\textbf{A}}(t)}{N S^{\textbf{A}}_{\textrm{E}_{\textrm{s}}}}\sim\eta (t)\;.
\end{equation}
Therefore, in a democratic system, we expect that for any two subsystems $\textbf{A}$ and $\textbf{B}$, the growth of entanglement entropies  satisfy the following simple scaling relation:
\begin{equation}\label{scaling}
S_{\textrm{E}}^{\textbf{A}}(t)\sim\frac{S^{\textbf{A}}_{\textrm{E}_{\textrm{s}}}}{S^{\textbf{B}}_{\textrm{E}_{\textrm{s}}}}S_{\textrm{E}}^{\textbf{B}}(t)\;,
\end{equation}
With the previous inequality we mean that during thermalization both quantities are only expected to be of the same order, but both stabilize at the same time. After stabilization both expressions evolve together.

The previous relation has a nice intuitive understanding. Consider a subsystem $\textbf{A}$ with `$a$'  degrees of freedom. Democracy of interactions means that the entanglement entropies associated to each of its degrees of freedom are all equal. But at the same time, democracy of interactions also ensure that this entanglement is due to correlations with $\textbf{A}$ and with its complement $\bar{\textbf{A}}$, so that equilibration of $\textbf{A}$ is ensured by the equilibration of its constituents degrees of freedom\footnote{There is a priori a certain tension between this result and the Brownian q-bit model presented in \cite{lashkari}. In that model it was studied a related quantity, called the purity of a subsystem. For a subsystem of size $k$ it is defined as $h_{k}=\textrm{Tr}(\rho_{k}^{2})$ . In a maximally mixed state, at stationarity, this purity is given by $h_{k}^{s}=2^{-k}$. In \cite{lashkari} it was found that the time evolution of this quantity in a certain democratic model was given at large times by $h_{k}=2^{-k}+a e^{-b t} k 2^{-k}$, where `$a$' and `$b$' are constants of $\mathcal{O}(1)$. In view of this, it was argued that the ratio between the scrambling time of large subsystems with small subsystems is of $\mathcal{O}(\log N)$, which seems incoherent with relation~(\ref{scaling}). But indeed a closer look to their result shows that it is perfectible compatible with our proposal if the time a small subsystem takes to thermalize is of $\mathcal{O}(\log N)$. Notice that their result can be written as $h_{k}(t)=h_{k'}(t)\frac{h_{k}^{s}}{h_{k'}^{s}}+a e^{-b t} (k-k') 2^{-k}$. The first term is analogous to the first term in~(\ref{scaling}), while the correction dies exponentially fast for times bigger than $\log N$, no matter the specific $k$. If the time it takes for a small subsystem to thermalize is bigger than $\log N$, then every subsystem thermalizes at the same time, no matter its size, supporting our generic argument, and pointing to a specific minimum time of $\mathcal{O}(\log N)$ for information scrambing. Also, reversing the argument, if their model stabilizes small subsystems in a certain $t\ll \log N$, the result challenges the fast scrambling conjecture, since unitarity implies that the lost information has to be found `somewhere', but democracy of interactions implies that it has to be found `everywhere', so that it is instantaneously scrambled. }.

Since for a non-equilibrium process with a thermal stationary state we have $S^{\textbf{A}}_{\textrm{E}_{\textrm{s}}}/S^{\textbf{B}}_{\textrm{E}_{\textrm{s}}}=a/b$, the previous scaling relation is just given in those cases by:
\begin{equation}\label{scalingT}
S_{\textrm{E}}^{\textbf{A}}(t)\sim\frac{a}{b}S_{\textrm{E}}^{\textbf{B}}(t)\;,
\end{equation}
a non trivial scaling relation which clearly does not hold for other type of systems, such as local models \footnote{Notice that relation~(\ref{scaling}) is valid through evolution at any time.}. In particular, the growth of entanglement entropy of one degree of freedom is related to the growth of entanglement entropy of half of the system by:
\begin{equation}\label{scaling2}
S_{\textrm{E}}^{\textbf{1}}(t)\sim\frac{2}{N}S_{\textrm{E}}^{\textbf{N}/2}(t)\:,
\end{equation}
a relation which imply the following peculiar property of democratic models:
\begin{itemize}
\item Relaxation of one subsystem (no matter its size) implies relaxation of every subsystem (no matter its size). In this precise sense, global relaxation/scrambling is equivalent to local relaxation.
\end{itemize}
This implies that scrambling in democratic models can be characterized by the breaking of the mean field approximation as time evolves from a factorized initial state. This approximation assumes that the global unitary evolution, acting on the full Hilbert space, is very close to a product of local unitaries, acting on each of the degrees of freedom. Mathematically it reads:
\begin{equation}
U_{\textrm{global}}(t)\simeq U_{\textrm{Mean-Field}}(t)=U_{\textbf{1}}(t)\otimes U_{\textbf{2}}(t)\otimes \cdots\otimes U_{\textbf{N}}(t)\;,
\end{equation}
where each $U_{\textbf{i}}(t)$ acts only in the reduced Hilbert space associated with the $i$'th degree of freedom. Beginning with a factorized state, $U_{\textrm{Mean-Field}}(t)$ cannot produce any type of entanglement, not even for a single degree of freedom, since it is a product of local unitaries. Therefore, the breaking of the mean field approximation can be characterized by the production of entanglement. But the produced entanglement has to be democratically spread over the system, due to the democratic structure of coupling constants in the theory, mathematically expressed through relation~(\ref{NoMarkov}). This connects with Ref. \cite{lashkari}, and we believe it completes the proof of fast scrambling for the last model the authors of \cite{lashkari} considered. Notice that the equivalence between the breaking of the mean field approximation and scrambling is only valid for democratic systems. For usual local systems, the mean field approximation can be broken without delocalizing the information associated to any degree of freedom. On other hand, for democratic systems, the breaking of the mean field approximation and information delocalization are unavoidably linked due to democracy and unitarity.

\subsection{Democratic walks and information fields}\label{sec42}

In this section we want to combine the democratic walks described in section~(\ref{sec2}) with the formalism of the information field. This section is therefore speculative, since we cannot prove that the evolution of the information field is given by one specific democratic walk \footnote{A knowledge of the exact evolution of the information field would prove or disprove the fast scrambling conjecture \cite{susskind}. We will comment more on this in section~(\ref{sec5}).}.

The physical situation is that described in section~(\ref{stoinfo}), a many-body quantum system with $N$ degrees of freedom and a factorized initial state
\begin{equation}
\vert \Psi_{t=0}\rangle =\vert\psi_{1}\rangle\otimes\vert\psi_{2}\rangle\otimes \cdots\otimes\vert\psi_{N}\rangle\;.
\end{equation} 
Given the previous initial quantum state, the initial state of the information field associated to subsystem $\textbf{1}$ is given by:
\begin{equation}
\mathcal{I}_{\textbf{1}}(\rho (t=0))=\left(\begin{array}{rcl}  &1 &\\&0 &\\ &\vdots &  \\ & 0 &
\end{array}\right)\;.
\end{equation} 
Since we know that unitary evolution of the initial quantum state will stochastically drive the information field to stationarity, we can use a random walk to model this evolution. The natural time scale associated to the discrete Markovian evolution is the time scale in which one interaction occurs, which is provided by the temperature, so that $n=t/\beta\equiv t_{\beta}$. Once an interaction occurs, $\mathcal{I}$ changes according to its specific kernel. Using the exact solution~(\ref{solution1}) of the democratic walk, the first relation we obtain reads:
\begin{equation}\label{evodensity}
\mathcal{I}_{\textbf{1}}(\rho (t_{\beta}))=\begin{pmatrix}
\alpha_{t_{\beta}} \\ \eta_{t_{\beta}} \\  \vdots \\ \eta_{t_{\beta}}
\end{pmatrix}\;,
\end{equation}
where
\begin{equation}\label{demosol}
\begin{pmatrix}
\alpha_{t_{\beta}} \\ \eta_{t_{\beta}}  
\end{pmatrix} =\begin{pmatrix}
\frac{1}{N}+\frac{(N-1)}{N}(\alpha-\eta)^{t_{\beta}} \\   \frac{1}{N}-\frac{(\alpha-\eta)^{t_{\beta}}}{N}
\end{pmatrix}\;.
\end{equation}
The structure of information leaking predicted by the democratic walk is equal to the one obtained by the generic considerations of the previous section. Besides, from the definition of $\mathcal{I}$ we obtain:
\begin{equation}\label{infolaw}
\frac{I(\textbf{1},\textbf{B}(t_{\beta}))}{2S_{\textrm{E}_{\textrm{s}}} N}= \eta_{t_{\beta}}=  \frac{1}{N}-\frac{(\alpha-\eta)^{t_{\beta}}}{N}\simeq \frac{1}{N}-\frac{\alpha^{t_{\beta}}}{N}\;,
\end{equation}
a relation from which we can read the evolution of entanglement entropy in these toy models:
\begin{equation}\label{entanglementlaw}
\frac{S_{\textrm{E}}^{\textbf{1}}(t_{\beta})}{S_{\textrm{E}_{\textrm{s}}}^{\textbf{1}}}=1-(\alpha-\eta)^{t_{\beta}}\simeq 1-\alpha^{t_{\beta}}\;.
\end{equation}
Notice that since $0<\alpha<1$, stochastic evolution drives entanglement entropy to stationarity at large enough times.

As in section~(\ref{fast}) we now consider different possible scalings of $\eta$. The interesting scalings are those in which there is a hierarchical separation between $\eta$ and $\alpha$. This is perfectly natural in the context of large $\mathcal{N}$ matrix models and other democratic models. Notice that when using a random walk to model information spreading, it is natural to assume that onsite kernel entries correspond to probabilities of free propagation in whatever basis scrambling is being considered. During free propagation there is no information transfer to other degrees of freedom, since there is no information transfer without the production of connected correlations, which are ultimately caused by microscopic interactions. On the other hand, off diagonal kernel entries are related to the strength of interactions, by which information can be shared through the production of connected correlations. It is well known that in large $\mathcal{N}$ matrix models, guage invariant operators can be constructed so as to be approximately free, up to $1/N$ corrections, where $N=\mathcal{N}^{2}$. In other words, operators propagate freely and do not share information up to $1/N$ corrections. This feature is consistently mirrored in the black hole geometric description, see \cite{tHooftS} and section~(\ref{sec5}), in which gravitational backreaction effects, responsible of producing connected correlations, are zero up to $\mathcal{O}(1/N)$ corrections. In turn, this is mirrored in our formalism by imposing a hierarchy between diagonal kernel entries and non-diagonal ones, as done in section~(\ref{sec2}). It is exactly in those cases in which interesting random walks appear.

\subsubsection{Walks with non-vanishing gap and quantum circuits}

Defining $0<\lambda_{\textrm{eff}}<1$, the first interesting class of random walks described in~(\ref{fast}) was the following:
\begin{equation}
\eta=\frac{\lambda_{\textrm{eff}}}{N} \sim\mathcal{O}(1/N)\;,
\end{equation}
and therefore
\begin{equation}
\alpha=1-(N-1)\,\eta=1-(N-1)\,\frac{\lambda_{\textrm{eff}}}{N}\;,
\end{equation}
In this case we had $0<\alpha<1$ in the thermodynamic limit, and so the second eigenvalue was bounded away from $1$, providing a spectral gap bounded away from zero, a feature that ensures fast scrambling of the democratic walk in the usual sense of stabilization of the stochastic process.

As we comment below, there are important drawbacks of this growth model. But it is interesting to notice that it is coherent with the quantum circuits models described in \cite{susskind}, based on previous works, see \cite{hayden} and references therein. These semi-democratic models are composed of $N$ spins. At each time we partition the system into $N/2$ couples at random, and we apply a random unitary matrix in the space of two spins to each couple. It is clear that for a single degree of freedom to affect the whole set of degrees of freedom we need to wait for at least $\log N$ steps, as described in $\cite{susskind}$. This is because the process of information diffusion proceeds in a tree like fashion \footnote{Although we are going to show that entanglement entropy evolution is similar in the quantum circuit model to~(\ref{entanglementlaw}) with $0<\alpha<1$ in the thermodynamic limit, the structure of information spreading in the quantum circuit, being tree like, is very different from the democratic one~(\ref{NoMarkov}). In this sense, these non-local random quantum circuits are not good models of truly democratic systems.}. But for these models, one further natural question which has not been considered in the past is for the evolution of entanglement entropy. In particular, is extensivity of entanglement entropy enough to characterize scrambling in these models? How much time does it take to get extensivity of entanglement entropy? To understand this evolution, consider any given number of time steps $k$, not scaling with the number of spins $N$. Also consider the initial state to be a product state. After $k$ steps we can ask for the graph representing the structure of interactions until that time. This {\it interaction graph} at time $k$ can be defined as follows. The vertices of this graph correspond to the different degrees of freedom of the system, while two vertices are connected by an edge if they have interacted by a random unitary. Aspects of time evolution can be observed through the evolving structure of this interaction graph. It is straightfoward to convince oneself that if the number of steps $k$ does not scale with $N$, the probability of two vertices having interacted more than once is negligible in the thermodynamic limit. We conclude that the resulting graph at time $k$ is just a random $k$-regular graph of $N$ vertices. Randomness here means the following: given the space of $k$-regular graphs (the collection of graphs in which each vertex has $k$ edges emanating from it) we associate the uniform measure to this collection, see \cite{avi1}. Given this observation, we now use a beautiful and profund result in combinatorics and graph theory, which states that random $k$-regular graphs are expander graphs \cite{fri}, see \cite{avi1} for a review. In turn, the defining notion of expander graphs is that the discrete notion of area is of the same order as the discrete notion of volume. In other words, if we take a subset of vertices, the number of them connected with one vertex outside the subset is of the same order as the total number of them. Given the results of \cite{us3}, this implies that entanglement entropies are already extensive after the first two steps in the random Q-bit model. We do not expect entanglement entropies to be a clear characterization of scrambling in these models, as in the case of expander graphs \cite{us3}. For these type of systems we expect an entanglement growth similar to~(\ref{entanglementlaw}), with a rapid convergence to extensivity, and a slow plateaux growth until the true stationarity has been reached \footnote{It would certainly be interesting to explore further the connection between these type of quantum circuits and random $k$-regular (expander) graphs.}. 

So from~(\ref{entanglementlaw}), in the case $0<\alpha<1$ in the thermodynamic limit, to measure the fast scrambling time scale from an entanglement entropy perspective we need to measure the change from $S_{\textrm{E}}^{\textbf{1}}=S_{\textrm{E}_{\textrm{s}}}^{\textbf{1}}\pm \mathcal{O}(1)$ at times of $\mathcal{O}(\beta)$ to $S_{\textrm{E}}^{\textbf{1}}=S_{\textrm{E}_{\textrm{s}}}^{\textbf{1}}\pm \mathcal{O}(1/N)$ at times of $\mathcal{O}(\beta \log N)$. From this perspective, the physics associated to the fast scrambling time scale is a little bit obscure, and seem to correspond to the physics in the so-called stationary plateaux.

\subsubsection{Walks with vanishing gap}\label{nonv}

Although the previous model has some nice features from various points of view, it has some drawbacks, as we now describe. As we have shown, $\eta=\frac{\lambda_{\textrm{eff}}}{N}$ implies that entanglement entropies are extensive after a time of $\mathcal{O}(\beta)$. Notice that this is true for any subsystem due to relations~(\ref{scaling}) and~(\ref{scaling2}). This feature, although natural from several perspectives, is also somewhat striking. Notice that for truly democratic systems, and opposed to the quantum circuits model described in the previous section, this feature implies that most of the information about the perturbation is already scrambled after a time of $\mathcal{O}(\beta)$, since any amount of information lost by the subsystem is spread over the whole set of degrees of freedom instantaneously, due to~(\ref{NoMarkov}). Since the scrambling time seems most naturally defined as the time in which an $\mathcal{O}(1)$ amount of information, with respect to the final stationary value $S_{\textrm{E}_{\textrm{s}}}$, is spread over an $\mathcal{O}(N)$ amount of degrees of freedom, we conclude that the previous model scrambles in $\mathcal{O}(\beta)$.

From a related perspective, the results \cite{lashkari,us3,stanford2014,sahakian, sahakian2,joansimon} point towards a different behavior of entanglement entropy for certain democratic models. Firstly, in Ref \cite{sahakian2} the authors find an exponential growth of entanglement entropy at initial times. They consider a subsystem with $N/2$ degrees of freedom, and their numerical findings read:
 \begin{equation}\label{saha}
 S_{\textrm{E}}^{\textbf{N}/2}(t)\sim e^{t/\beta}\:,
 \end{equation}
which is of $\mathcal{O}(N)$ by times of $\mathcal{O}(\beta \log N)$. Using relation~(\ref{scaling}), the exponential growth of entanglement entropy found in \cite{sahakian2} turns into $S_{\textrm{E}}^{\textbf{1}}(t)\sim e^{t/\beta}/N$ for a single degree of freedom, a result reminiscent of the one found in \cite{lashkari} concerning the breaking of the mean field approximation in democratic models. This is consistent with the generic arguments given in the previous section concerning the implications of democratic interactions in mutual information and entanglement entropies, summarized by relations~(\ref{NoMarkov}) and~(\ref{scaling}), which seems to imply that for democratic systems, the computation of the time in which the mean field approximation breaks down is equivalent to the computation of the scrambling time.


Let us see what scaling of $\eta$ leads qualitatively to their results. Given~(\ref{entanglementlaw}), the entanglement entropy of a democratic system as a function of $\alpha=1-(N-1)\,\eta$ is given by:
\begin{equation}
\frac{S_{\textrm{E}}^{\textbf{A}}(t_{\beta})}{S_{\textrm{E}_{\textrm{s}}}^{\textbf{A}}}=1-(\alpha-\eta)^{t_{\beta}}\simeq 1-\alpha^{t_{\beta}}\;.
\end{equation}
To find the value of $\eta$ providing extensivity of entanglement entropy at times of $\mathcal{O}(\beta \log N)$ and not earlier, notice that if $\eta$ dies with $N$ faster than $1/N$, then we can approximate $\log \alpha\simeq -(N-1)\,\eta$, and we arrive to:
\begin{equation}\label{equivalenceent}
\frac{S_{\textrm{E}}^{\textbf{A}}(t_{\beta})}{S_{\textrm{E}_{\textrm{s}}}^{\textbf{A}}}=1-\alpha^{t_{\beta}}\simeq 1-e^{-(N-1)\,\eta \, t_{\beta}}
\end{equation}
Extensivity of entanglement entropy is obtained at times of $\mathcal{O}(\frac{1}{\eta\, N})$. Therefore, with $\eta\sim \frac{1}{N\log N}$ extensivity of entanglement entropy is exactly obtained at times of $\mathcal{O}(\beta\log N)$, and not before. This is the second class of democratic models studied in~(\ref{fast}). Interestingly, for this value of $\eta$ the production of entanglement entropy is suppressed in the thermodynamic limit at times of $\mathcal{O}(\beta)$:
\begin{equation}\label{equivalenceent2}
\frac{S_{\textrm{E}}^{\textbf{A}}(\beta)}{S_{\textrm{E}_{\textrm{s}}}^{\textbf{A}}}\simeq \frac{\beta}{\log N}\;.
\end{equation}
This is an interesting relation, since it means that the perturbation is not entangled in the semiclassical lmit, or thermodynamic limit $N\rightarrow\infty$, until times of order of the scrambling time $\mathcal{O}(\beta\log N)$, i.e until the perturbation reaches the streched horizon, see \cite{us1,us2} and section~(\ref{sec5}) below.

This heirarchy seems also consistent with the results found in \cite{stanford2014,joansimon}, in the context of holographic CFT's with large central charge. In those studies entanglement production with the environment was also suppressed until times of order the scrambling time, as in our simpler model. The advantage of our model is that it makes the global properties of information spreading more transparent.

The hierarchy is also consistent with previous comments, regarding relation~(\ref{scaling}) and the results in \cite{lashkari}. In this model the relaxation of a small subsystem takes $\mathcal{O}(\log N)$, and so all subsystems can scramble in this time as well, without contradicting the results of \cite{lashkari}.

Finally, through the AdS/CFT correspondence, there is a simple way to estimate the time it will take for a perturbation to get entangled due to interactions with the environment. We compute it in section~(\ref{sec5}), following the results in \cite{us3} together with the framework developed in \cite{vijayent}.

\subsection{On the fate of information in the thermodynamic limit}\label{sec43}

Relations~(\ref{uni}) and~(\ref{unitarity}) convey the explicit relation between the stochastic evolution of the information field and unitarity  of the underlying microscopic theory. The breaking of one of them implies the breaking of the other. Assuming that democratic walks are good models of the stochastic evolution of $\mathcal{I}$, it turns out to be an interesting question to analyze the thermodynamic/large $N$ limit of the model.

One might be tempted to think that the thermodynamic limit is conceptually similar to that of other random walks, for example those walks associated to euclidean or expander graphs, which are discrete versions of diffusion processes defined over flat and hyperbolic spaces respectively, see \cite{us3}. In this section we show that they are radically different. The conclusions might have certain relevance on the issue of information loss in black hole physics.

Let us first discuss the thermodynamic limit of random walks defined over $k$-regular graphs. Euclidean nets and expander graphs fall naturally into this class of graphs. In this $k$-regular case, the $N\rightarrow\infty$ limit just affects the time scales associated with the walk, sending them to infinity. It affects `the size' of the transition matrix, but otherwise leaves this matrix `intact', since the entries are independent of $N$. The limit implies that the initially non-zero entry of the probability distribution decays monotically to zero without bound. This last feature is usually taken to mean information loss in the process. Information loss is here associated to the infinite heat reservoir that is produced when taking the $N\rightarrow\infty$ limit.  

The last conclusion, conerning information loss, is slightly misleading. At any time $t$ in the process, the information that leaked out from the initially non-zero entry of the probability distribution can be found in a region with a radial size $\Omega_{\textrm{spread}}(t)$. This function $\Omega_{\textrm{spread}}(t)$ depends on the specific $k$-regular graph considered, but it does not scale with $N$ in the thermodynamic limit. The conclusion might be expressed in two different ways. At fixed $t$, in the large $N$ limit the following properties hold:
\begin{itemize}
\item $\Omega_{\textrm{spread}}$ is of $\mathcal{O}(1)$ for $k$-regular graphs.
\item Verifying conservation of probability in the stochastic process requires to look only through $\mathcal{O}(1)$ entries of the probability/density distribution.
\end{itemize}
Antoher related and important aspect of the limit is the following:
\begin{itemize}
\item Stochasticity of the transition matrix is preserved through the limit. The limit just makes the matrix bigger, but it does not affect any particular entry of the matrix.
\end{itemize}


Interestingly, the previous statements are violated by the democratic walks. In the first case, for which the hierarchy reads:
\begin{equation}
\eta=\frac{\lambda_{\textrm{eff}}}{N}\sim \mathcal{O}(1/N)\;,
\end{equation}
and
\begin{equation}
\alpha=1-(N-1)\,\eta=1-(N-1)\,\frac{\lambda_{\textrm{eff}}}{N}\;,
\end{equation}
we have that in the thermodynamic limit, the transition matrix $M$ is a diagonal matrix with digaonal entries given by $\alpha$:
\begin{equation}
M\rightarrow M_{\infty}=\textrm{Diag}\,\left( \alpha,\alpha, \cdots, \alpha\right) \:.
\end{equation}
Therefore the evolution of the probability distribution is given by:
\begin{equation}\label{decay}
M_{\infty}^{t_{\beta}}\begin{pmatrix}
1 \\ 0 \\ 0 \\ \vdots \\ 0
\end{pmatrix}=\begin{pmatrix}
\alpha^{t_{\beta}} \\ 0 \\ 0 \\ \vdots \\ 0
\end{pmatrix}\;.
\end{equation}
The initially non zero entry decays exponentially fast, but without any visible information leaking in the thermodynamic limit. This can be summarized by:
\begin{itemize}
\item Stochasticity of the transition matrix is lost when taking the limit. In this limit the transition matrix is a diagonal matrix with entries between cero and one, so it does not conserve probability through evolution. Given~(\ref{unitarity}) and~(\ref{uni}), this must then be supported by a breaking of unitarity of the underlying democratic quantum model.
\end{itemize}
This leads to the following speculative scenario. In Matrix models, taking the thermodynamic limit amounts to tracing out the non-planar sector of the theory.  A `tracing out/integrating out' procedure always leaves us with an effective theory for the reduced subsystem which is non-unitary to some extent. This is due to the neglected interactions between the subsystem studied and the environment. At high energies, in the black hole sector,  this non-unitarity might be expected to be strong, since the planar sector is highly entangled with the non-planar sector. The planar sector would then be described by a superoperator \footnote{A Superoperator is the most generic evolution map within quantum mechanics. It is just constrained by conservation of the normalization of the quantum state, but not by the conservsation of its von Neumann entropy. Superoperators appear naturally when considering reduced evolution laws, associated to reduced subsystems, in a global unitary evolving 
 quantum system. They are therefore natural in the context of open quantum systems. We refer to the nice notes \cite{preskillnotes} for more details.}.
Since the Einsten-Hilbert action is the dual description of the planar sector, the old Hawking's alternative presented in \cite{paradox} could then be well acomodated within AdS/CFT, or other conjectured dualities with Hamiltonians having a large number of degrees of freedom $N$ with democratic interactions. The decay of the perturbation survives the thermodynamic limit, being the colletive $\mathcal{O}(1)$ effect of $\mathcal{O}(N)$ softly interacting degrees of freedom\footnote{This model share some resemblance and might of interest for the model presented in \cite{cesar}}, but the correlations are killed in this limit, and the leaked information is lost. This effect is summarized by relation~(\ref{decay}).

In the second case, for which the hierarchy reads:
\begin{equation}
\eta=\frac{1}{N\log N}\;,
\end{equation}
and
\begin{equation}
\alpha=1-(N-1)\,\eta\;,
\end{equation}
we have that in the thermodynamic limit $M$ becomes the identity matrix:
\begin{equation}
M\rightarrow M_{\infty}=\textrm{Diag}\,\left( 1,1, \cdots, 1\right) \:.
\end{equation}
Therefore there is no evolution of the probability distribution:
\begin{equation}\label{decayI}
M_{\infty}^{t_{\beta}}\begin{pmatrix}
1 \\ 0 \\ 0 \\ \vdots \\ 0
\end{pmatrix}=\begin{pmatrix}
1 \\ 0 \\ 0 \\ \vdots \\ 0
\end{pmatrix}\;.
\end{equation}
In this case, we conclude that in the large $N$ limit there is simply no decay. This then relates to the discussion in the previous section, in which we showed that entanglement entropy is suppressed at initial times in the thermodynamic limit. In this situation there is simply no spread of information.

\section{Black hole applications: free fall, entanglement and scrambling}\label{sec5}

Relations~(\ref{NoMarkov}) and~(\ref{scaling}) signal the extremely peculiar, but otherwise simple structure of information flows in democratic systems. In particular, relation~(\ref{scaling}) implies that stabilization of a small subsystem implies stabilization of the whole system as well. This has a promising perspective, since it means that we can study scrambling in democratic systems by studying subsystems of single degrees of freedoms alone.

In this regard there is a clear instance in wich we can study such subsystems. This is in the context of matrix models \cite{matrix} and AdS/CFT \cite{adscft}. As it is well known, high energy states in these theories are dual to black holes in the geometric description. In this geometric description it is very natural to consider perturbations consisting on infalling particles. We can estimate generically the time by wich they become entangled with black hole degrees of freedom by joining the ideas developed in \cite{us3} to compute the near horizon running coupling constants together with the results of \cite{vijayent}, regarding the perturbative expansion of entanglement entropy in weak coupling scenarios. This estimate will again point towards the second random walk, described in~(\ref{nonv}), as the good toy model of information spreading in black hole systems.

First, from \cite{vijayent} we see that having a certain degree of freedom interacting with an environment through an interaction coupling $\lambda\ll 1$, perturbation theory tells us that the entanglement entropy is
\begin{equation}\label{ent}
S_{\textrm{E}}\propto \lambda^{2}\log\lambda^{2}\;,
\end{equation}
at lowest order in $\lambda$. The previous relation is simple to understand heuristically. The amplitude for an intercation is of $\mathcal{O}(\lambda)$, so the probability is given by $\mathcal{O}(\lambda^{2})$. Relation~(\ref{ent}) is just the Shannon entropy of the probability of an interaction, valid for small $\lambda$

With this in mind we just need to estimate the coupling with the Hawking radiation for an infalling particle. Using standard renormalization group ideas, for an energy quanta of energy $E$ the effective dimensionless coupling is given by
\begin{equation}
\lambda_{\textrm{eff}}\propto (E/m_{\textrm{p}})^{\alpha}\;,
\end{equation}
where the exponent might depend on the dimension of the operator considered and it will be irrelevant for the argument. For a typical infalling quanta of energy $T$, this is then given by
\begin{equation}
\lambda_{\textrm{eff}}\propto (T/m_{\textrm{p}})^{\alpha}\;.
\end{equation}
Since the coupling is suppressed by inverse powers of $m_{\textrm{p}}$, it might seem that there is no production of entanglement due to formula~(\ref{ent}). The crucial aspect to notice is that for an infalling particle the energy is not a constant, since it is blushifted due to the background geometry. This blueshift is universal, see \cite{us1,us3,stanford2014,bound}, and it is given by \footnote{We want to remark that this aspect was first highlighted in \cite{us3}  in the present context of near horizon running couplings, within the optical representation of the dynamics, while later in \cite{stanford2014} this universal energy blueshift was nicely connected to entanglement evolution in the dual field theory. Finally, very recently, it has been used in \cite{bound} to argue for a universal bound on the growth of chaos in many body quantum systems.}
\begin{equation}\label{lambdaevo}
\lambda_{\textrm{eff}}(t)\propto (T e^{\frac{2\pi}{\beta}t}/m_{\textrm{p}})^{\alpha}\;.
\end{equation}
The entanglement of the perturbation is therefore
\begin{equation}\label{entp}
S_{\textrm{E}}\propto \lambda^{2}(t)\log\lambda^{2}(t)\;,
\end{equation}
becoming of $\mathcal{O}(1)$ when the proper temperature becomes Planckian $T e^{\frac{2\pi}{\beta}t}\sim m_{\textrm{p}}$, i,e when the probability of an interaction is of $\mathcal{O}(1)$. This is the location of the brick wall \cite{tHooftS} or streched horizon \cite{complementarity}, which for a generic black hole in $d+2$ dimensions implies that the time by which the pertubation becomes fully entangled is given by
\begin{equation}\label{tent}
t_{\textrm{ent}}=\frac{\beta}{2\pi d}\log S_{*}\sim \beta \log S_{*}\;,
\end{equation}
where $S_{*}$ is the entropy of a horizon patch of size $\beta$, a so-called thermal cell \footnote{If the horizon has a volume $V$ it is simply defined by $
S_{\textrm{BH}}=VT^{d} S_{*}$, where $S_{\textrm{BH}}$ is the entropy of the full black hole}, see \cite{us1} for a derivation of the previous relation. Notice that the previous time-scale is not precisely the time scale found in \cite{stanford2014,joansimon}, since here we are interested in the time-scale in which the perturbation gets fully entangled with black hole degrees of freedom, and not when the perturbation has enough energy to distort the black hole geometry. In \cite{stanford2014,joansimon} the perturbations they consider are able to fall way past the streched horizon, affecting the geometry when their proper mass is of the order of the total black hole mass, and not when it is just of Planckian order.

Since the interactions are democratic, this entanglement production is democratically spread/scrambled by the same time~(\ref{tent}). Relation~(\ref{entp}) points to a connection betweent the near horizon coupling constants and entanglement production in democratic systems. In this context everything can be characterized by the evolution of perturbations consisting of single degrees of freedom. For example, to read off the Lyapunov exponent of the near horizon region, which is given by $\lambda_{\textrm{Lyapunov}}=\frac{2\pi}{\beta}$, see \cite{us1,bound}, it should be enough to compute the entanglement entropy of single degrees of fredom in democratic systems.

\section{Concluding remarks}

From an abstract perspective, random walks and more generically stochastic processes are arguably the simplest models of information spreading in many-body systems, whether classical or quantum. In the form of Fokker-Planck type equations, they help us to understand the coarse grained transport properties of locally conserved charges in translationally invariant systems, which otherwise are very difficult to study from an exact microscopic description. Not surprisingly, the Hamiltonians which are expected to describe black holes are special in this regard, see~(\ref{nonlocal}). Being completely non-local, neither there is a clear space in which to diffuse, nor is there some clear physical quantity spreading over the system. On the other hand, it is interesting and reasonable to expect that there are simple equations describing information transport for these systems as well.

In this vein the purpose of this article has been twofold. From one side, given the non-local structure of matrix models~(\ref{nonlocal}), we have studied the properties of random walks defined on weighted complete graphs, see~(\ref{rel1}),~(\ref{rel2}) and Fig~(\ref{demo}). These democratic walks turn out to be explicitly solvable, giving a transparent image of information spreading in democratic systems, see~(\ref{solution1}),~(\ref{alet}) and~(\ref{solution2}) for the explicit solution.  It was further shown that there is a window range of parameters which fit well with the fast scrambling conjecture \cite{susskind}, providing a new family of systems showing fast scrambling behavior.

From the other side we have proved that stochastic processes are rigorous models of quantum thermalization. The construction is based on the developments presented in \cite{uscod}, and goes through the definition of appropriate information fields $\mathcal{I}$. These information fields can be unambiguously computed from the exact quantum state, a feature implying that the microscopic unitary evolution generates a well defined time evolution for these fields as well. Two crucial properties were shown to hold for the evolution of $\mathcal{I}$. The first aspect is stochasticity of the induced evolution. More precisely, the induced evolution conserves normalization of the information field if and only if the underlying evolution is unitary. In this scenario, somewhat surprisingly, unitarity and stochasticity are transparently connected to each other. This is appealing since unitarity conserves the Von Neumann entropy of the quantum state, while stochasticity will not generically conserve the Shannon entropy of the information field as time evolves. The second crucial aspect of the construction concerns the \emph{typical} information field $\mathcal{I}$, i.e the $\mathcal{I}$ associated to a random state in the Hilbert space. Given the results of $\cite{uscod}$, this is shown to be the stationary distribution, given by $\mathcal{I}^{n}=1/N$, where $n=1, \cdots, N$ runs over the different degrees of freedom of the system. The conclusion is that non-equilibrium unitary processes, in which an atypical quantum state approaches typicality, are mirrored by stochastic evolution of the information field towards stationarity. This is summarized by relations~(\ref{randomuni}) and~(\ref{sto}). This approach deserves more development, since it is a novel and promising way to study chaotic phenomena, being classical or quantum, from a unified information theoretic perspective. In particular, subtle notions of information scrambling and chaos in many body quantum mechanics might be studied through the mixing properties of the classical stochastic evolution of the information field. It would then be extremely interesting to explore other related definitions of these information fields, and extend the framework to QFT.

In the last part we applied both constructions and studied information spreading in democratic systems. The two generic features we find, which are closely related to each other, are relations~(\ref{NoMarkov}) and~(\ref{scaling}), both expressing the simple structure of information spreading in democratic systems. In these type of systems any information leaking is instantaneously spread over the whole set of degrees of freedom, and in this precise sense it is instantaneously scrambled. Therefore we just need to agree on how much information we want the chosen subsystem to lose, and compute the time for this specific leaking. This feature can be summarized by stating that, for democratic systems, \emph{global} thermalization/scrambling is equivalent to \emph{local} thermalization/relaxation of the perturbation. Assuming that the properties of information flows do not strongly depend of its initial location, this feature was shown to have an interesting implication, given by~(\ref{scaling}), which states that the equlibration of entanglement entropy of a given subsystem, no matter its size, implies the equilibration of the whole system as well \footnote{This result is consistent with the results presented in \cite{stanford2014,joansimon}. It is also consistent with the results of \cite{lashkari} if the relaxation time of small subsystems is of $\mathcal{O}(\log N)$, a feature consistent with the second model studied in~(\ref{sec2}), and with the estimate done in~(\ref{sec5}).}. In section~(\ref{sec4}) we argued that this implies that scrambling in democratic systems is equivalent to the breaking of the mean field approximation.

There is an instance, particularly interesting for the present concerns, in which such an entanglement entropy can be estimated. That is the evolution of an infalling particle through the near horizon region of a black hole. We estimate this in section~(\ref{sec5}), by joining the results unravelled in \cite{us3}, regarding the behavior of the near horizon running coupling constants, with the framework of \cite{vijayent}, regarding the computation of entanglement entropies in weak coupling scenarios. The result is that the particle becomes entangled by the time it reaches the streched horizon, and not before, a feature which is consistent with the fast scrambling conjecture, see~(\ref{lambdaevo}) and~(\ref{tent}). Due to democracy of interactions, no further scrambling is needed at the streched horizon. From this perspective, the scrambling time is just seen as the time in which the probability of an interaction with the Hawking radiation is of $\mathcal{O}(1)$. We conclude that to read off the properties of the near horizon, the running couplings for example, it is enough to study the behavior of entanglement entropy for single degrees of freedom in democratic systems.

Subsections~(\ref{sec42}) and~(\ref{sec43}) are speculative. We presented them because of their interest, because they can be connected with other works on the field, and because they motivate further research in this direction. Assuming a Markov approximation for the stochastic evolution of $\mathcal{I}$, and using the democratic walk model as a kernel for the information field, all the previous generic considerations are seen to follow in a straightforward way. Besides, having the specific exact solutions, given by~(\ref{solution1}) and~(\ref{alet}), we commented on the implications of each of the two `fast scrambling' democratic walks unravelled in section~(\ref{sec2}) on the behavior of entanglement entropy. The first democratic walk seems incoherent with previous results, in particular with the democratic brownian model presented in \cite{lashkari}. But interestingly, for one specific hierarchy of parameteres we are able to qualitatively reproduce the results found in \cite{lashkari,stanford2014,sahakian2,joansimon}, together with the estimate carried out in section~(\ref{sec5}), while at the same time giving a unified perspective of entanglement growth and entanglement spreading in democratic systems. Finally, in section~(\ref{sec43}) we made some comments on the thermodynamic limit of the model. This limit turns out to be extremely different from the analoge limit in the context of local models, a result which sheds new light on the fate of information, and in particular on the mechanisms of information loss, in black hole physics.

The reasons for these subsections to be speculative is that we cannot prove that a given democratic walk, with a pre-specified hierarchy of transitions matrix entries, furnishes the appropriate kernel for the evolution of the information field. But we hope to come to this problem in the future, and that the construction presented motivate different computations and contribute to unify different approaches in this field, since it connects spreading of correlations and evolution of entanglement entropy in a transparent way.

\newpage
\bigskip{\bf Acknowledgements:}
We want to express a most sincere gratitude to Jose L. Barbon, for many years sharing his knowledge about physics and in particular black hole physics. We are also specially grateful to Simone Paganeli for many discussions concerning information theory and thermalization, and Phil Szepietowski for useful comments on the manuscript. We also want to thank Vijay Balasubraminan, Panos Betzios, Riccardo Bursato, Marcos Crichigno, Nava Gaddam, Umut Gursoy, Vadim Oganesyan, Pasquale Sodano, Amilcar Queiroz, Gerard t'Hooft and Stefan Vandoren for useful discussions, the IIP (Natal) for funding the first stages of this project and setting a free and warm environment for independent research and CUNY university and its department "Initiative for Theoretical Physics" for hospitality in the spring of 2014, during which part of this work was developed. This research has been funded by the Delta-Institute for Theoretical Phyics (D-ITP), which is funded by the Dutch Ministry of Education, Culture and Science (OCW).

\appendix

\newpage


\end{document}